\documentclass[10pt,journal,compsoc]{./IEEEtran}

\title{
Requirements of API Documentation:\\
A Case Study into Computer Vision Services
}

\author{Alex Cummaudo,
    Rajesh Vasa,
    John Grundy,
    and Mohamed Abdelrazek%
    \IEEEcompsocitemizethanks{
        \IEEEcompsocthanksitem A. Cummaudo and R. Vasa are with the Applied Artificial Intelligence Institute, Deakin University, 1 Gheringhap St, Geelong Victoria 3220, Australia. E-mail: \{ca, rajesh.vasa\}@deakin.edu.au.
        \IEEEcompsocthanksitem J. Grundy is with the Faculty of Information Technology, Monash University, Wellington Rd, Clayton Victoria 3800, Australia. E-mail: john.grundy@monash.edu.
        \IEEEcompsocthanksitem M. Abdelrazek is with the School of Information Technology, Deakin University, 1 Gheringhap St, Geelong Victoria 3220, Australia. E-mail: mohamed.abdelrazek@deakin.edu.au.
    }
}
\usepackage[utf8]{inputenc} 
\usepackage[hyphens]{url}   
\usepackage[table]{xcolor}  
\usepackage[normalem]{ulem} 
\usepackage{emptypage}      
\usepackage{blindtext}      
\usepackage{ragged2e}       
\usepackage{enumitem}       
\usepackage{subcaption}     
\usepackage{pdflscape}      
\usepackage{multicol}       
\usepackage{gensymb}        
\usepackage{bold-extra}     
\usepackage{amsmath}        
\usepackage{amssymb}        
\usepackage{enumitem}       
\usepackage{flushend}       

\usepackage[multiple]{footmisc}
\newcommand{\footnoteurl}[2]{\footnote{{#1} last accessed #2.}}

\usepackage{booktabs}
\usepackage{multirow}
\usepackage{makecell}

\usepackage{longtable}

\usepackage{graphicx}
\graphicspath{{images/}{../images/}}

\usepackage{bibentry}
\usepackage[numbers,compress]{natbib}
\usepackage[labeled,resetlabels]{multibib}
\newcites{W}{\ }
\newcites{S}{\ }
\newcommand{\citepweb}[1]{\mbox{\citeW[W\hspace{-.75ex}][]{#1}}}

\usepackage{framed,mdframed}
\usepackage{tablefootnote}

\usepackage[nameinlink]{cleveref}
\crefname{appendix}{appendix}{appendices}
\usepackage{fontawesome}
\def\circlenotpresent{\faCircleO}
\def\circlepartialpresent{\faAdjust}
\def\circlepresent{\faCircle}
\newcommand{\dimcat}[1]{\textsc{\small[\textbf{#1}]}}

\def \dima{Descriptions of API Usage}
\def \dimb{Descriptions of Design Rationale}
\def \dimc{Descriptions of Domain Concepts}
\def \dimd{Existence of Support Artefacts}
\def \dime{Overall Presentation of Documentation}

\def\SurveyParticipantsTotal{104}
\def\SurveyParticipantsInternal{22}
\def\SurveyParticipantsInternalResponseRate{50.00\%}
\def\SurveyParticipantsExternalSnowball{38}
\def\SurveyParticipantsExternalMTurk{44}
\def\SurveyParticipantsExternalTotal{82}
\def\SurveyParticipantsExternalPartialResponses{13}

\def\SurveyParticipantsExternalPartialResponsesCompletionRate{43.23\%}

\def\SurveyParticipantsFullResponses{91}

\def\SurveyParticipantsExternalMTurkRejected{12}
\def\SurveyParticipantsExternalResponseTooSmall{37}
\def\SurveyParticipantsTotalReached{153}
\def\SurveyParticipantsTotalResponseRate{67.97\%}

\def\dimcatILSvalueAFive{0.71}

\begin{document}

\IEEEtitleabstractindextext{\begin{abstract}
Using cloud-based computer vision services is gaining traction, where developers access AI-powered components through familiar RESTful APIs, not needing to orchestrate large training and inference infrastructures or curate/label training datasets.
However, while these APIs \textit{seem} familiar to use, their non-deterministic run-time behaviour and evolution is not adequately  communicated to developers.
Therefore, improving these services' API documentation is paramount---more extensive documentation facilitates the development process of intelligent software.
In a prior study, we extracted 34 API documentation artefacts from 21 seminal works, devising a taxonomy of five key requirements to produce quality API documentation. We extend this study in two ways. Firstly, by surveying \SurveyParticipantsTotal{} developers of varying experience to understand what API documentation artefacts are of \textit{most value} to practitioners. Secondly, identifying which of these highly-valued artefacts are or are not well-documented through a case study in the emerging computer vision service domain.
We identify: (i)~several gaps in the software engineering literature, where aspects of API documentation understanding is/is not extensively investigated; and (ii)~where industry vendors (in contrast) document artefacts to better serve their end-developers.
We provide a set of recommendations to enhance intelligent software documentation for both vendors and the wider research community.
\end{abstract}
}

\maketitle

\IEEEdisplaynontitleabstractindextext
\IEEEpeerreviewmaketitle

\IEEEraisesectionheading{\section{Introduction}\label{sec:introduction}}

\IEEEPARstart{I}{mproving} API documentation quality is a valuable task for any API. Succinct API documentation of good quality facilitates productivity \citep{Lethbridge:2005jv,myersstylos2016,7503516}, and therefore improved quality is better engineered into a system \citep{mcleod2011factors}. Where application developers integrate new services into their systems via APIs, their productivity is affected either by inadequate skills (\textit{``I've never used an API like this, so must learn from scratch''}) or, where their skills are adequate, an imbalanced cognitive load that causes excessive context switching (\textit{``I have the skills for this, but am confused or misunderstand''}).
As a real-world use case, consider intelligent computer vision services, in which an AI-based component produces a non-deterministic result based on a machine-learnt data-driven algorithm, rather than a predictable, rule-driven one \citep{Cummaudo:2019icsme}. These services use machine intelligence to make predictions on images such as object labelling or facial recognition \citepweb{GoogleCloud:Home,AWS:Home,Azure:Home,IBM:Home,Pixlab:Home,Clarifai:Home,Cloudsight:Home,DeepAI:Home,Imagaa:Home,Talkwaler:Home,Megvii:Home,TupuTech:Home,YiTuTech:Home,SenseTime:Home,DeepGlint:Home}. The impacts of poor and incomplete documentation results in developer complaints on online discussion forums such as Stack Overflow \citep{Cummaudo:2020icse}. Many comments show that developers do not think in the non-deterministic mental model of the designers who created the computer vision services. They ask many varied questions from their peers to try and clarify their understanding.

It is therefore important to ensure developers have access to high-quality API documentation artefacts when consuming these services. Vendors should cover all documentation artefacts that the wider developer community find valuable, and the research community should aide in this process by investigating with types of information that comprise these artefacts, or the aspects of information design to best present this information.
What causes a developer to be confused when using an API, and how to mitigate it via improved documentation, has been largely explored by researchers for \textit{conventional} APIs (an overview is provided in \cref{tse2020:sec:related-work}). Various studies provide a myriad of recommendations into the value of API documentation artefacts based on both qualitative and quantitative analyses, involving developer opinions (from surveys), observation of developers, event logging or content analysis (see \cref{tse2020:fig:sms}). Such guidelines propose ways for developers, managers, and solution architects can construct systems better with improved documentation. 

However, there does not yet exist a consolidated \textit{systematic} review of this literature. Further, few studies offer a taxonomy to consolidate these guidelines together, and there still lacks a consolidated effort to capture guidelines on the requirements of good quality API documentation. Studies that produce these guidelines from literature are largely scattered across multiple sources. Investigating the ways by which these guidelines are produced can provide software engineering researchers with better insight into the research methods and data collection techniques used to produce these guidelines. Some studies, for example, use case studies, others use focus groups and brainstorming, or interviews and surveys. The extent to which researchers rely on developer opinion for API documentation guidelines is evident, and gaps in the methodological approaches that researchers use should be emphasised to shine light into new ways of conducting research in this important area. Furthermore, systematically capturing the information distilled from these guidelines into a readily accessible, consolidated taxonomy (designed to assist writing API documentation) must be validated in real-world circumstances to assess its efficacy with practitioners.

In our prior work, we proposed an API documentation taxonomy that was comprised of 21 key primary sources \citep{Cummaudo:2019esem}. This paper significantly extends our previous work by addressing limitations in the existing taxonomy, thus refining it. Previously, we developed a metric for each dimension (topmost-layers) and category (leaf nodes) within the taxonomy \citep{Cummaudo:2019esem}. This metric is an indication of the specific areas of API documentation software engineering researchers have focused their efforts, as measured by the ratio of papers that investigated or reported various issues concerning the documentation artefacts defined within our taxonomy. For the context of this paper, we refer to this metric as an `in-literature' score, or ILS. Within this paper, we build upon this facet but \textit{in-practice} by assessing the efficacy of our taxonomy against developers using a survey instrument inspired by the System Usability Scale (SUS) \citep{Brooke:1996ua}. Each artefact within the taxonomy is measured against this instrument for its utility, and a metric is produced to indicate how well developers \textit{value} each of these artefacts. We refer to this metric as an `in-practice' score, or IPS. (Details for how the IPS is calculated are in given in \cref{tse2020:sec:validation:survey:analysis}.) We then identify the artefacts that are highly researched, the ones that developers demand the most, and where gaps in these artefacts remain for future research exploration.

Lastly, while our prior work focused on \textit{generalised} API documentation, in this extension, we apply our taxonomy to a case study of interest: i.e., better documenting computer vision services. We empirically assess the taxonomy against three popular computer vision services, namely Google Cloud Vision \citepweb{GoogleCloud:Home}, Amazon Rekognition \citepweb{AWS:Home} and Azure Computer Vision \citepweb{Azure:Home}. For each category in our taxonomy, we assess whether the respective service's documentation contains, partially-contains or does not contain the documentation artefact from our taxonomy, thus determining the extent to which the requirements of good API documentation are met within the vendors' own documentation.
From this, we triangulate each ILS and IPS value against the service's level of inclusion of its respective documentation artefact, thereby making a judgement as to where the services can improve their documentation to make them more complete. Lastly, we present a ranking of each artefact for where research or vendors should be focus their documentation efforts that is of high value to both developers \textit{and} to industry vendors.

Thus, through this triangulation of the taxonomy with existing literature, utility to practitioners, and application via a case study (computer vision services), we summarise three aspects of API documentation by identifying:
\begin{enumerate}[label=(\roman*)]
    \item the documentation artefacts that been extensively studied by researchers, and those that warrant further attention by the software engineering research community (via high/low ILS values);
    \item the documentation artefacts that are considered to be the most- and least-important from a practitioner's point of view (via high/low IPS values);
    \item the documentation artefacts that have been well-established by vendors (via our case study on three prominent computer vision services). 
\end{enumerate}

To demonstrate how our taxonomy was developed, we include an extended revision of the systematic mapping study (SMS) from our existing work. The taxonomy we proposed consists of five key requirements: (1)~\dima{}; (2)~\dimb{}; (3)~\dimc{}; (4)~\dimd{}; and (5)~\dime{}. Following this, we developed a survey instrument to assess the overall utility of each of the artefacts that contribute towards these five requirements, which consisted of 43 questions of alternating positive and negative sentiment. 
We then narrow our focus down to our case study by applying the prioritised documentation artefacts (as identified by the survey) to three computer vision services. Once our surveys were complete, we provide some general guidelines as to where cloud computer vision services can make improvements to their API documentation. Lastly, we compare and contrast the results from our SMS to the results of the survey and of our case study, thereby identifying where future research efforts into API documentation should focus to give the biggest value back to practitioners.

Our key contributions in this work are:

\begin{itemize}
  \item a score metric for each category that indicates where the highest research priorities have been in the existing literature;
  \item a score metric assessing the efficacy of the 34 categories that empirically reflects what artefacts are of the highest value from a \textit{practitioner} point of view;
  \item a heuristic validation of each artefact against computer vision services, assessing where existing computer vision service API documentation needs improvement;
  \item a number of practical recommendations for computer vision service vendors to better improve the quality of their API documentation; and
  \item an identification of the gaps for future research into API documentation based on the highest need by developers but, so far, has captured the least attention by researchers.
  
\end{itemize}
This paper is structured as follows: \cref{tse2020:sec:related-work} presents related work; \cref{tse2020:sec:method} is divided into two subsections, the first describing how primary sources were selected in the SMS with the second describing the development of our taxonomy from these sources; \cref{tse2020:sec:findings} presents the taxonomy; \cref{tse2020:sec:validation} describes how we developed a survey instrument of 43 questions to validate the taxonomy against developers, and assess its efficacy against the three popular computer vision services selected; \cref{tse2020:sec:tax-analysis} presents the findings from our validation analysis; \cref{tse2020:sec:limitations} describes the threats to validity of this work; and \cref{tse2020:sec:conclusions} provides concluding remarks and the future directions of this study. Additional materials are provided in \cref{tse2020:tab:taxonomy,tse2020:tab:docsources,tse2020:sec:primary-sources,tse2020:sec:online-artefacts,tse2020:sec:suggested-improvements,tse2020:sec:survey}; referenced online artefacts are prefixed with `W' and can be found in \cref{tse2020:sec:online-artefacts}.

\section{Related Work}
\label{tse2020:sec:related-work}

\subsection{Systematic Reviews in Software Documentation}

Systematic reviews into how developers produce and use software documentation gives researchers consolidated insights into the efforts of multiple, disparate API documentation studies. For example, a recent \citeyear{Nybom:2018ef} study explored 36 API documentation generation tools and approaches, and analysed the tools developed and their inputs and documentation outputs \citep{Nybom:2018ef}. The findings from this study emphasise that the largest effort in API documentation tooling is to assist developers to generate either example code snippets and/or templates or natural language descriptions of the API directly from the program's source code. These snippets or descriptions can then be placed in the API documentation, thereby increasing the efficiency at which API documentation can be written. Additionally, tools from 12 studies target the maintainability of existing APIs of existing APIs, while tools from 11 studies target the correctness and accuracy of the documentation by validating that what is written in the documentation is accurate to the technical structure of the API. From the end-developer's perspective, some tools (17 studies) help target improvements to the developer's understandability and learnability of new APIs by linking in examples directly with questions such as on Stack Overflow. 
However, the results from this study regards the \textit{tooling} used to either assist in producing, validating or learning from API documentation. While this is a systematic study with key insights into the types of tooling produced, there is still a gap for a SMS in what \textit{guidelines} have been produced by the literature in developing natural language documentation itself---and how well developers \textit{agree} to those guidelines---which our work has addressed.

An extensive SMS into studies presented in the \textit{overall} software documentation domain was given in \citet{ZHI2015175}. This study reviewed a set of 69 papers from 1971 to 2011 to develop a systematic map on the various research aspects relating to documentation cost, benefit and quality, finding that 38\% of papers propose novel techniques while 29\% contribute empirical evidence (i.e., validation and evaluation papers---see \cref{tse2020:sec:data-extraction}). The authors find that a majority of papers discuss quality aspects of software documentation, namely the quality attributes of completeness, consistency and accessibility, and that the main usage of software documentation regards maintenance aid and program comprehension. Another key insight---relevant to our study---found that, on average, survey-based studies into documentation involved 106 participants and generally these participants were from the same (or only two)  organisations. However, unlike our study, this study formalises the documentation efforts of \textit{any} software document, and not exclusively into API documentation artefacts required to help developers produce software. Further, our study differs in that the results from our study are consolidated into a structured taxonomy, instead of a meta-model which \citeauthor{ZHI2015175} perform, which is then triangulated against a real-world use case (i.e., intelligent computer vision services) and software developers via a survey.

\subsection{API Usability and Documentation Knowledge}

API usability and its impact on documentation knowledge is an imperative area of study, since it provides useful links between API documentation and more technical issues related to API design or tools. Extensive discussions from \citet{myersstylos2016} and \mbox{\citet{7503516}} encapsulate a 30-year effort to evaluate and improve API usability through lenses adapted human-computer interaction research. Essentially, by treating a developer as the `end-user' of an API (i.e., interacting and programming with the API in their own systems), the authors discuss various case studies by which API usability was improved by various human-centred approaches, resulting in improved learnability of the API in addition to improved productivity and effectiveness in using the API. While the methods are primarily used for end-user usability testing, their observations highlight the importance of good aesthetic and interaction design of developer's tooling and the need for new tooling to augment what developers already do to reduce learning overhead. An extensive review of the usability methods used, and their benefits to API usability, demonstrates how various techniques---grounded through established usability guidelines and frameworks---can be used to assess how an API's usability impacts its key stakeholders (i.e., API designers, developers, and end-users). The role of API \textit{documentation} in context to an API's overall usability is imperative; for instance, limited documentation on a particular API (and limited code snippets) is often a key complaint to poor API usability \citep{myersstylos2016}. Exploring aspects on information design elements within API documentation is therefore critical to mitigate such complaints.

In \citet{Watson:2012uy}, the authors performed a heuristic assessment from 35 popular APIs against 11 high-level universal design elements of API documentation. Of these 35 APIs, 28 were open-source software repositories and seven came from commercial independent software vendors. Two coders manually inspected each API's respective documentation sets, starting from the documentation's entry page and using the navigation features of the documentation to further explore the documentation. Both coders evaluated each of the 11 heuristics, noting whether they could be found. This study highlighted how many APIs, even popular ones, fail to grasp these basic design elements. For example, 25\% of the documentation sets did not provide any basic overview documentation to the API. Therefore, from a practitioner's perspective, the study describes a high-level overview of how certain documentation artefacts address their needs and whether they are typically found in documentation. However, while the methodological approach used in this study to assess the heuristics is similar to our approach, the heuristics themselves used within \citeauthor{Watson:2012uy}'s study is based on only three seminal works and only contains 11 design elements. Our study extends these heuristics and structures them into a consolidated, hierarchical taxonomy which we then validate against practitioners.

A taxonomy of distinct knowledge patterns within reference documentation by \citet{Maalej2013} classified 12 distinct knowledge types. Unlike our work, which uses a SMS of existing studies as the source of our taxonomy development, this study uses a grounded method via theoretical sampling of the API documentation of two mature (extensively documented) open source systems. This was performed by each author to elicit a list of knowledge types over an iterative six month process. The taxonomy was then evaluated against the JDK 6 and .NET 4.0 frameworks using a sample of 5574 documentation units and 17 trained coders to assign each knowledge type to the documentation unit. Results showed that the functionality and structure of these APIs are well-communicated, although core concepts and rationale about the API are quite rarer to see. The authors also identified low-value `non-information'---described as documentation that provides uninformative boilerplate text with no insight into the API at all---which was  substantially present in the documentation of methods and fields in the two frameworks. They recommend that developers factor their 12 distinct knowledge types into the process of code documentation, thereby preventing low-value non-information, and thus developers can use the patterns of knowledge to evaluate the content, organisation, and utility of their own documentation. The development of their taxonomy consisted of questions to model knowledge and information, thereby capturing the reason about disparate information units independent to context; a key difference to this paper is the \textit{systematic} taxonomy approach utilised and the source of information of our taxonomy (i.e., existing literature).

\subsection{Computer Vision Services}

Recent studies into cloud-based computer vision services have demonstrated that poor reliability and robustness in computer vision can `leak' into end-applications if such aspects are not sufficiently appreciated by developers. A study by \citet{Hosseini:2018jr} showed that Google Cloud Vision's labelling fails when as little as 10\% noise is added to the image. Facial recognition classifiers are easily confused by modifying pixels of a face and using transfer learning to adapt one person's face into another \citep{Wang:2018vl}. Our own prior work found that the non-deterministic evolution of these types of services is not adequately communicated to developers \citep{Cummaudo:2019icsme}, resulting in lost developer productivity whereby developers ask fundamental questions about the concepts behind these services, how they work, and where better documentation can be found \citep{Cummaudo:2020icse}. This paper continues this line of research by providing a means for service providers to better document their services using a taxonomy and suggested improvements.

\section{Taxonomy Development}
\label{tse2020:sec:method}

\begin{figure*}[t]
\centering
\includegraphics[width=.8\linewidth]{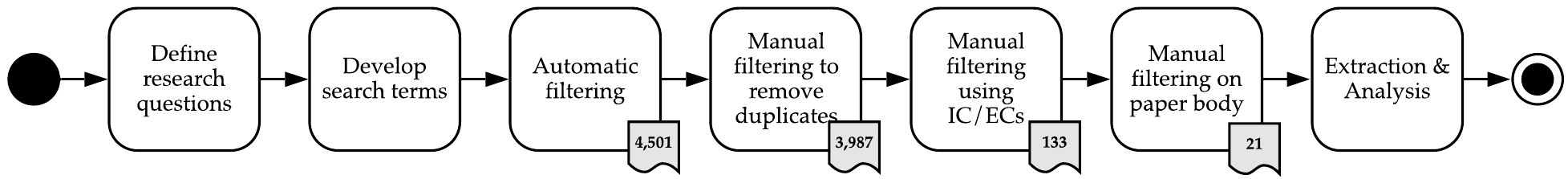}
\caption[Filtering steps used in the systematic mapping study]{A high level overview of the filtering steps from defining and executing our search query to the data extraction of our primary studies. Number of accepted papers resulting from each filtering step is shown.}
\label{tse2020:fig:filtering}
\end{figure*}

We developed our taxonomy under two primary phases. First, we conducted a SMS identifying API documentation studies, following guidelines by \citet{Kitchenham:2007dd} and \citet{Petersen:2008td} (\cref{tse2020:sec:method:lit-review}). A high level overview of this first phase is given in \cref{tse2020:fig:filtering}. Second, we followed a software engineering taxonomy development method by \citet{Usman:2017hn} (\cref{tse2020:sec:method:taxonomy-development}) based on the findings of our SMS, which involved an extensive validation involving real-world developers and contextualised with computer vision APIs (\cref{tse2020:sec:validation}).

\subsection{Systematic Mapping Study}
\label{tse2020:sec:method:lit-review}

\begin{figure*}[t]
  \includegraphics[width=\linewidth]{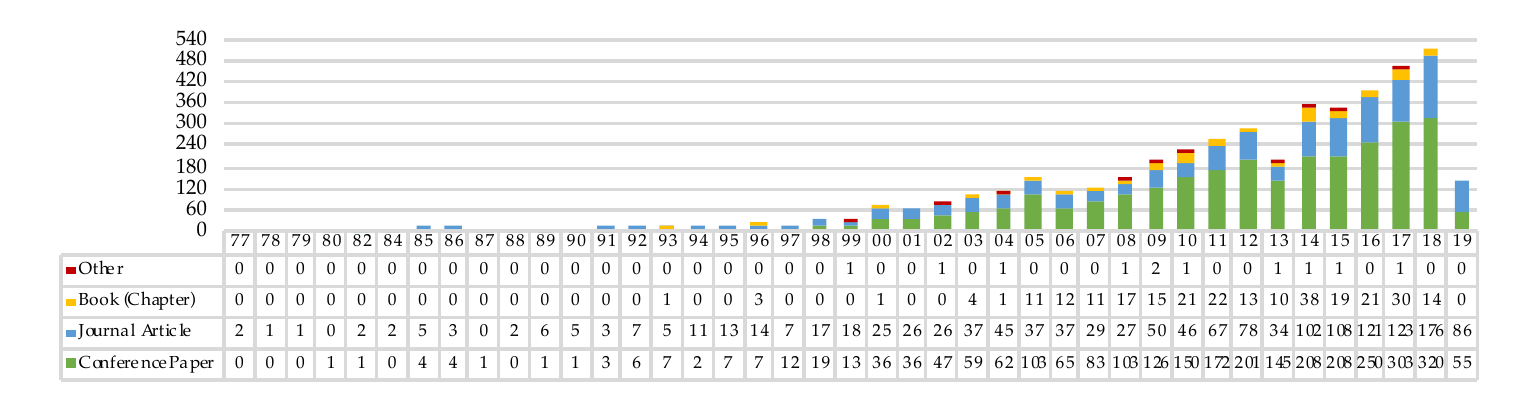}
  \caption[SMS search results, by years]{Search results by year and venue type.}
  \label{tse2020:fig:slr-years}
\end{figure*}

\subsubsection{Research Questions (RQs)}

The first step in producing our SMS was to pose two RQs:
\begin{itemize}[leftmargin=\parindent]
  \item \textbf{RQ1:} What documentation `knowledge' do API documentation studies contribute?
  \item \textbf{RQ2:} How is API documentation studied?
\end{itemize}
Our intent behind RQ1 was to collect as many studies provided by literature on how API documentation should be written using natural language, i.e., not using assistive tooling. In this regard, documentation `knowledge' encompasses any natural language API documentation artefact associated with the implementation of an application using a third-party API. As the goals of this study are to arrive at a taxonomy encapsulating the requirements of good API documentation (\cref{tse2020:sec:findings}), we sought to arrive at studies that provide useful information to developers that informs the relevance and value of which aspects of API documentation are more useful than others. This captures the knowledge that developers need to know about what aspects of their APIs should be documented and the artefacts by which they do this. This helped us shape and form the taxonomy provided in \cref{tse2020:sec:findings}. Secondly, RQ2's intent was to understand how the studies derive at their conclusions, thereby helping us identify gaps in literature where future studies can potentially focus.

\subsubsection{Automatic Filtering}

As done in similar software engineering studies \citep{Glass:2002wa,Usman:2017hn,GAROUSI2019101}, we explored  automatic filtering of online databases. We defined which SWEBOK  knowledge areas \citep{IEEE:1990wp} were relevant to devise a search query. Our search query was built using related knowledge areas, relevant synonyms, and the term `software engineering' (for comprehensiveness) all joined with the OR operator. Due to the lack of a standard definition of an API, we include the terms: `API' and its expanded term; software library, component and framework; and lastly SDK and its expanded term. These too were joined with the OR operator, appended with an AND. Lastly, the term `documentation' was appended with an AND.
Our final search string was:

\begin{framed}
\noindent
\parbox{\linewidth}{
\scriptsize
( ``software design'' \textbf{OR} ``software architecture" \textbf{OR} ``software construction" \textbf{OR} ``software development" \textbf{OR} ``software maintenance" \textbf{OR} ``software engineering process" \textbf{OR} ``software process" \textbf{OR} ``software lifecycle" \textbf{OR} ``software methods" \textbf{OR} ``software quality" \textbf{OR} ``software engineering professional practice" \textbf{OR} ``software engineering" ) \textbf{AND} ( API \textbf{OR} ``application programming interface" \textbf{OR} ``software library" \textbf{OR} ``software component" \textbf{OR} ``software framework" \textbf{OR} sdk \textbf{OR} ``software development kit" ) \textbf{AND} ( documentation )
}
\end{framed}

We executed the query on all available metadata (title, abstract and keywords) in May 2019 against Web of Science\footnoteurl{\url{http://apps.webofknowledge.com}}{23 May 2019}  (WoS), Compendex/Inspec\footnoteurl{\url{http://www.engineeringvillage.com}}{23 May 2019} (C/I) and Scopus\footnoteurl{\url{http://www.scopus.com}}{23 May 2019}. We selected three particular primary sources given their relevance in software engineering literature (containing the IEEE, ACM, Springer and Elsevier databases) and their ability to support advanced queries \citep{Brereton:2007by,Kitchenham:2007dd}. A total 4,501 results\footnote{Raw results can be located at \url{http://bit.ly/2KxBLs4}.} were found, with 549 being duplicates. \Cref{tse2020:tab:search-results} displays our results in further detail (duplicates not omitted); \cref{tse2020:fig:slr-years} shows an exponential trend of API documentation publications produced within the last two decades. (As this search was conducted in May 2019, results taper in 2019.)

\begin{table}[tb]
  \caption[Summary of search results in API documentation]{Search results and publication types}
  \label{tse2020:tab:search-results}
  \centering
  \begin{tabular}{l|lll|l}
    \toprule
    \textbf{Publication type} &
    \textbf{WoS} &
    \textbf{C/I} &
    \textbf{Scopus} &
    \textbf{Total} \\
    \midrule
    Conference Paper & 27 & 442 & 2353 & 2822 \\
    Journal Article & 41 & 127 & 1236 & 1404\\
    Book & 23 & 17 & 224 & 264\\
    Other & 0 & 5 & 6 & 11\\
    \midrule
    \textbf{Total} & 91 & 591 & 3819 & 4501\\
    \bottomrule
  \end{tabular}
\end{table}

\subsubsection{Manual Filtering}

A follow-up manual filtering stage followed the 4,501 results obtained by automatic filtering. As described below, we applied the following inclusion criteria (IC) and exclusion criteria (EC) to each result:

\begin{enumerate}[leftmargin=2\parindent,label=\textbf{IC\arabic*}]
  \item Studies must be relevant to API documentation: specifically, we exclude studies that deal with improving the technical API usability (e.g., improved usage patterns);
  \item Studies must discuss artefacts that document APIs;
  \item Studies must be relevant to software engineering as defined in SWEBOK;
\end{enumerate}
\begin{enumerate}[leftmargin=2\parindent,label=\textbf{EC\arabic*}]
  \item Studies where full-text is not accessible through standard institutional databases; 
  \item Studies that do not propose or extend how to improve the official, natural language documentation of an API;
  \item Studies proposing a third-party tool to enhance existing documentation or generate new documentation using data mining (i.e., not proposing strategies to improve official documentation);
  \item Studies not written in English;
  \item Studies not peer-reviewed.
\end{enumerate}
\smallskip

Each of these ICs and ECs were applied to every paper  after exporting all  metadata of our results to a spreadsheet. The first author then curated the publications using the following revision process.

Firstly, we read the publication source---to rapidly omit non-software engineering papers---as well as the author keywords, title, and abstract of all 4,501 studies. As some studies were duplicated between our three primary sources, we needed to remove any repetitions. We sorted and reviewed any duplicate DOIs and fuzzy-matched all very similar titles (i.e., changes due to punctuation between primary sources), thereby retaining only one copy of the paper from a single database. Similarly, as there was no limit do our date ranges, some studies were republished in various venues (i.e., same title but different DOIs). These were also removed using fuzzy-matching on the title, and the first instance of the paper's publication was retained. This second phase resulted in 3,987 papers.

Secondly, we applied our inclusion and exclusion criteria to each of the 3,987 papers by reading the abstract. Where there was any doubt in applying the criteria to the abstract alone, we automatically shortlisted the study. We rejected 427 studies that were unrelated to software engineering, 3,235 were not directly related to documenting APIs (e.g., to enhance coding techniques that improve the overall developer usability of the API), 182 proposed new tools to enhance API documentation or used machine learning to mine developer's discussion of APIs, and 10 were not in English. This resulted in 133 studies being shortlisted to the final phase.

Thirdly, we re-evaluated each shortlisted paper by re-reading the abstract, the introduction and conclusion. We removed a further 64 studies that were on API usability or non API-related  documentation (i.e., code commenting). At this stage, we decided to refine our exclusion criteria to better match the research goals of this study by including the word `natural language' documentation in EC2. This removed studies where the focus was to improve technical documentation of APIs such as data types and communication schemas. Additionally, we removed 26 studies as they were related to introducing new tools (EC3), 3 were focused on tools to mine API documentation, 7 studies where no guidelines were provided, 2 further duplicate studies, and a further 10 studies where the full text was not available, not peer reviewed or in English. Books are commonly not peer-reviewed (EC5), however no books were shortlisted within these results. This final stage resulted in 21 primary studies for further analysis, and the mapping of primary study identifiers to references S1--21 can be found in \cref{tse2020:sec:primary-sources}.

As a final phase, we conducted reliability analysis of our shortlisting method. We conducted intra-rater reliability of our 133 shortlisted papers using the test-retest approach suggested by \citet{Kitchenham:2007dd}. We re-evaluated a random sample of 10\% of the 133 shortlisted papers a week after initial studies were shortlisted. This resulted in \textit{substantial agreement}~\citep{Landis:1977kv}, measured using Cohen's kappa ($\kappa=0.7547$).

\subsubsection{Data Extraction \& Systematic Mapping}
\label{tse2020:sec:data-extraction}

Of the 21 primary studies, we conducted abstract key-wording adhering to \citeauthor{Petersen:2008td}'s guidelines \citep{Petersen:2008td} to develop a classification scheme.
An initial set of keywords were applied for each paper in terms of their methodologies and research approaches (RQ2), based on an existing classification schema used in the requirements engineering field by \citet{Wieringa:2006vd}. These are: \textit{evaluation papers}, which evaluates existing techniques currently used in-practice; \textit{validation papers}, which investigates proposed techniques not yet implemented in-practice; \textit{experience papers}, which are written by practitioners in the field and provide insight into their experiences of adopting existing techniques; and \textit{philosophical papers}, which presents new conceptual frameworks that describes a language by which we can describes our observations of existing or new techniques, thereby implying a new viewpoint for understanding phenomena. For example, documenting APIs using code snippets is a commonly used practice by developers (see the primary sources listed in \cref{tse2020:tab:taxonomy}), and conducting an experiment exploring how quickly practitioners achieve this would be an evaluation paper. In contrast, a validation paper explores novel techniques that are proposed but not yet implemented in practice; for example, a paper proposing that APIs should document success stories so that developers know where, why, and how the API was successfully implemented may test this novel technique via field study experiments (e.g., interviewing developers on the new technique) without reference to real-world examples. A paper written by a group of developers sharing their insights into the improvements of their documentation before and after providing extensive tutorials would be an experience paper. Philosophical papers may propose entirely new vocabulary to explore API documentation, devising new frameworks from which other researchers can explore the field from a new viewpoint.

After all primary studies had been assigned keywords, we noticed that all papers used field study techniques, and thus we consolidated these keywords using \citeauthor{Singer:2007tu}'s framework of software engineering field study techniques \citep{Singer:2007tu}. \citeauthor{Singer:2007tu} captures both study techniques \textit{and} methods to collect data within the one framework, namely: \textit{direct techniques}, including brainstorming and focus groups, interviews and questionnaires, conceptual modelling, work diaries, think-aloud sessions, shadowing and observation, participant observation; \textit{indirect techniques}, including instrumenting systems, fly-on-the-wall; and \textit{independent techniques}, including analysis of work databases, tool use logs, documentation analysis, and static and dynamic analysis. 

\Cref{tse2020:tab:extraction} describes our data extraction form, which was used to collect relevant data from each paper. \Cref{tse2020:fig:sms} presents our systematic mapping, where each study is mapped to one (or more, if applicable) of methodologies plotted against \citeauthor{Wieringa:2006vd}'s research approaches. We find that a majority of these studies survey developers using direct techniques (i.e., interviews and questionnaires) and some performing structured documentation analysis. Few studies report recent experiences; literature reports the artefacts that document APIs from evaluation research, in addition to some validation studies. There are few experience papers describing anecdotal evidence, and almost no philosophical papers that describe new conceptual ways at approaching API documentation as a large majority of existing work either evaluates existing (in-practice) strategies or validates the effectiveness of new strategies.

\begin{table}[tb]
  \caption[Data extraction form used for systematic mapping study]{Data extraction form}
  \label{tse2020:tab:extraction}
  \centering
  \begin{tabular}{l|p{0.6\linewidth}}
    \toprule
    \textbf{Data item(s)} &
    \textbf{Description}
    \\
    \midrule
    Citation metadata & Title, author(s), years, publication venue, publication type \\
    Artefact(s) discussed & As per IC2, the study must identify at least one API documentation artefact \\
    Evaluation method & Did the authors evaluate their proposed artefacts? If so, how? \\
    Primary technique & The primary technique used to devise the artefact(s) \\ 
    Secondary technique & As above, if a second study was conducted \\
    Tertiary technique & As above, if a third study was conducted \\
    Research type & The research type employed in the study as defined by \citeauthor{Wieringa:2006vd}'s taxonomy \\
    \bottomrule
  \end{tabular}
\end{table}

\subsection{Development of the Taxonomy}
\label{tse2020:sec:method:taxonomy-development}

A majority of taxonomies produced in software engineering studies are often made extemporaneously \citep{Usman:2017hn}. For this reason, we decided to proceed with a systematic approach to develop our taxonomy using the guidelines provided by \citet{Usman:2017hn}, which are extended from lessons learned in more mature domains. In this subsection, we outline the 4 phases and 13 steps taken to develop our taxonomy based on \citeauthor{Usman:2017hn}'s technique. \citeauthor{Usman:2017hn}'s final \textit{validation} phase is largely detailed within \cref{tse2020:sec:validation} after we present our taxonomy in \cref{tse2020:sec:findings}.

Formally, \citeauthor{Usman:2017hn} provides guidelines to define these units under the first six stages under the planning phase. In our study, our preliminary phase involves answering the following:

\begin{figure}[t!]
  \centering
  \includegraphics[width=\linewidth]{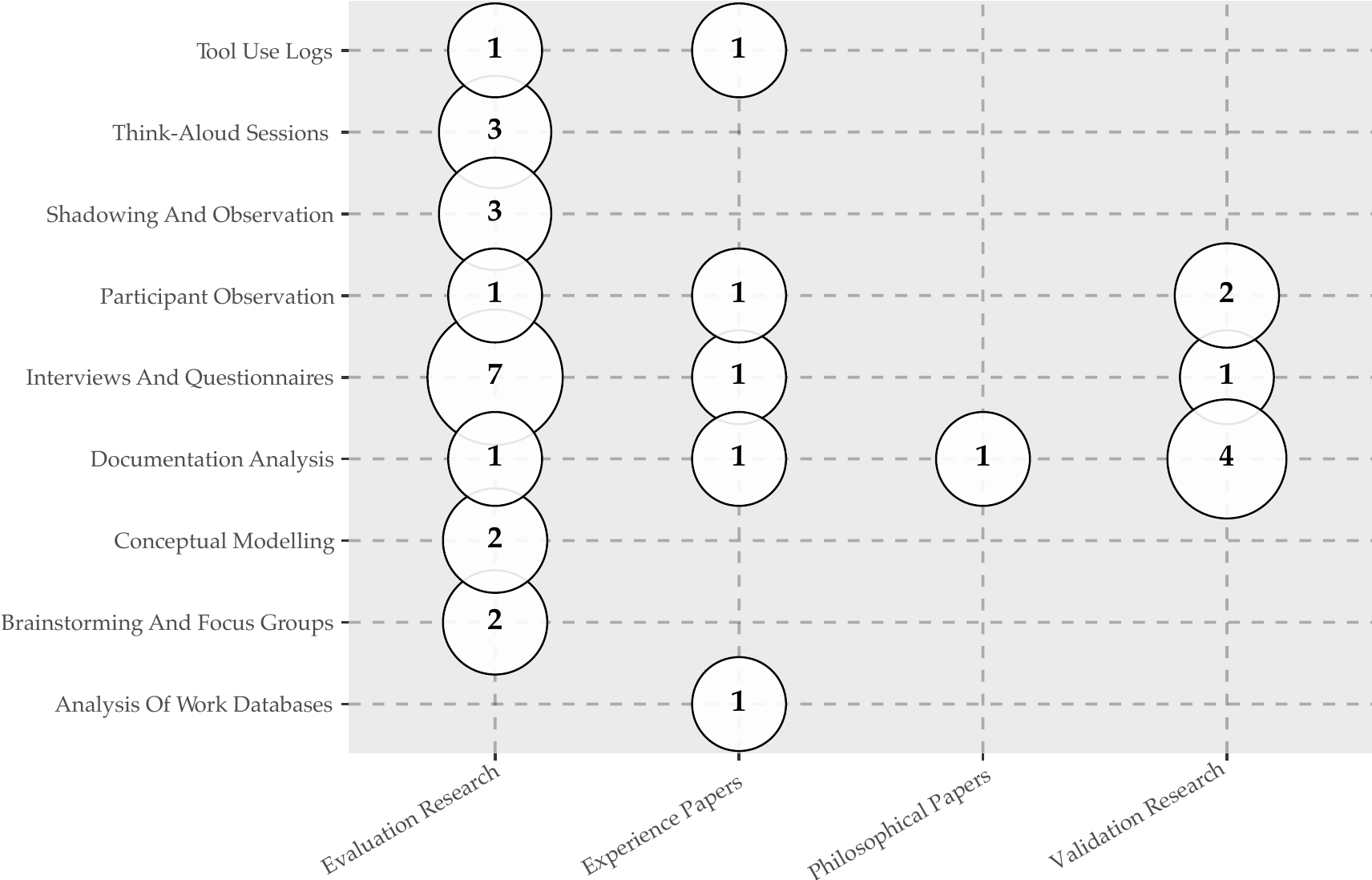}
  \caption[A systematic map of API documentation studies]{Systematic map: field study technique vs research type}
  \label{tse2020:fig:sms}
\end{figure}

\begin{enumerate}[label=\textbf{(\arabic*)}]
  \item \textit{define the software engineering knowledge area}: The software engineering knowledge area, as defined by the SWEBOK, is software construction;
  \item \textit{define the objective}: The main objective of the proposed taxonomy is to define a set of categories that enables to classify different facets of natural language API \textit{documentation} artefacts (not API \textit{usability}) as reported in existing literature;
  \item \textit{define the subject matter}: The subject matter of our proposed taxonomy is  documentation artefacts of APIs;
  \item \textit{define the classification structure}: The classification structure of our  proposed taxonomy is \textit{hierarchical};
  \item \textit{define the classification procedure}: The procedure used to classify the documentation artefacts is qualitative; 
  \item \textit{define the data sources}: The basis of the taxonomy is derived from field study techniques (see \cref{tse2020:sec:data-extraction}).
\end{enumerate}

\subsubsection{Identification and extraction phase} The second phase of the taxonomy development involves \textbf{(7)}~\textit{extracting all terms and concepts} from relevant literature, which we have achieved from our SMS. These terms are then consolidated by \textbf{(8)}~\textit{performing terminology control}, as some terms may refer to different concepts and vice-versa. For example, \citeauthor{Watson:2012uy} defines one of the heuristics used in the study's experiment as ``sample apps to understand how to use the elements of an API in context and as another source from which to copy program code... a sample app is a complete application that includes examples of the API as well as the other functions that comprise a complete program''~\citep{Watson:2012uy}. In this case, the term `sample app', `program code', and `complete application' were extracted as a term of interest and noted. Similarly, in \citet{Robillard:2009uk}, the phrase `applications' is used to define a category of example code snippets which ``consists of code segments from complete applications'' and is generally some form of ``demonstration samples sometimes distributed with an API... that developers can download from various source code repositories''~\citep{Robillard:2009uk}. Again, the phrase `complete applications', `demonstration samples', `download', and `source code' was identified as a terms of interest and noted. Once all papers were read, we consolidated a list of all of these noted highlights to help consolidate the terms and perform terminology control. In this example, the phrase `Downloadable source code demonstrating complete sample applications' was consolidated from both \citeauthor{Watson:2012uy} and \citeauthor{Robillard:2009uk}'s studies, which---in addition to the other primary studies that iteratively changed wording slightly due to steps (9--10)---formed the basis of the taxonomy dimension \dimcat{A7}.

\subsubsection{Design phase} \label{tse2020:sec:method:taxonomy-development:design-phase} The design phase identified the core dimensions and categories within the extracted data items. The first step is to \textbf{(9)}~\textit{identify and define taxonomy dimensions}; for this study we utilised a bottom-up approach to identify each dimension, i.e., extracting the categories first and then nominating which dimensions these categories fit into using an iterative approach. As we used a bottom-up approach, step (9) also encompassed the second stage of the design phase, which is to \textbf{(10)}~\textit{identify and describe the categories} of each dimension. Thirdly, we \textbf{(11)}~\textit{identify and describe relationships} between dimensions and categories, which can be skipped if the relationships are too close together, as is the case of our grouping technique which allows for new dimensions and categories to be added. The last step in this phase is to \textbf{(12)}~\textit{define guidelines for using and updating the taxonomy}. The taxonomy is as simple as a checklist that can be heuristically applied to API documentation, and each dimension is malleable and covers a broad spectrum of artefacts; while we do not anticipate any further dimensions to be added, new categories can easily be fitted into one of the dimensions (see \cref{tse2020:sec:conclusions}). We provide guidelines for use in our application of the taxonomy against computer vision services within \cref{tse2020:sec:findings,tse2020:sec:tax-analysis}.

\subsubsection{Validation phase} In the final phase of taxonomy development, taxonomy designers must \textbf{(13)}~\textit{validate the taxonomy} to assess its usefulness. \citet{Usman:2017hn} describe three approaches to validate taxonomies: (i)~orthogonal demonstration, in which the taxonomy's orthogonality is demonstrated against the dimensions and categories, (ii)~benchmarking the taxonomy against similar classification schemes, or (iii)~utility demonstration by applying the taxonomy heuristically against subject-matter examples. In our study, we adopt utility demonstration by use of a survey and heuristic application of the taxonomy against real-world case-studies (i.e., within the domain of computer vision services). This is is discussed in greater detail within \cref{tse2020:sec:validation}.

\section{A Taxonomy for API Documentation}
\label{tse2020:sec:findings}

\begin{figure}[h!]
\centering
\includegraphics[width=.9\linewidth]{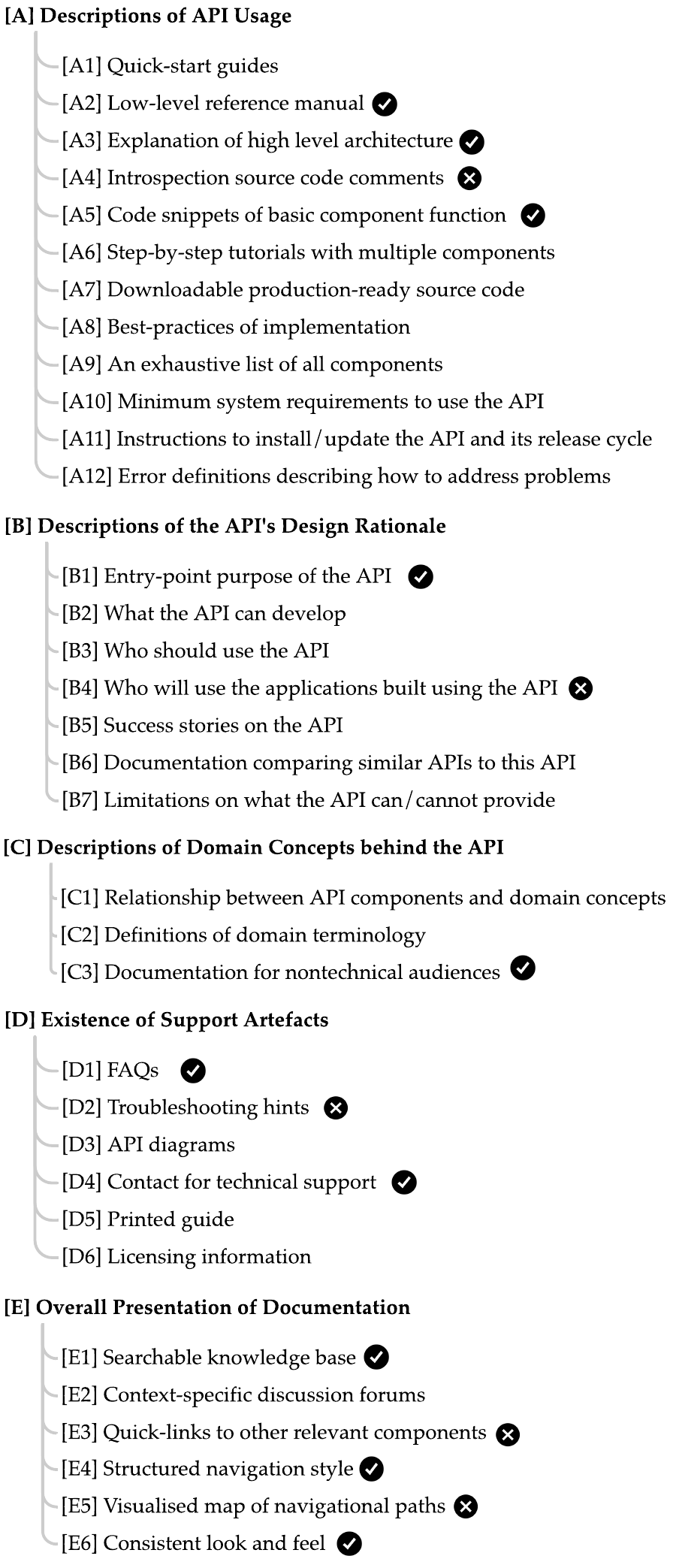}
\caption[Our proposed API documentation taxonomy]{Our proposed taxonomy: The requirements of good-quality API documentation (dimensions) represented through individual documentation artefacts (categories).}
\label{tse2020:fig:taxonomy}
\end{figure}

Our taxonomy consists of five dimensions (labelled A--E). These five dimensions are made of 34 categories, which represent  API documentation artefacts that contribute towards these dimensions. In the context of our taxonomy, a category can represent (i)~discrete and self-contained documentation artefacts (e.g., quick start guides \dimcat{A1}), (ii)~additional information used to describe the API (e.g., licensing information about the API \dimcat{D6}), or (iii)~aspects regarding the information design of this documentation (e.g., consistent look and feel \dimcat{E6}). \textbf{Collectively, the categories form the \textit{requirements} of good quality API documentation, as expressed through the five dimensions.} When worded as questions, each dimension respectively covers the following:

\begin{itemize}
  \item \dimcat{A}~\textbf{\dima{}}: \textit{how} does the developer use this API for their intended use case?
  \item \dimcat{B}~\textbf{\dimb{}}: \textit{when} should the developer choose this particular API for their intended use case?
  \item \dimcat{C}~\textbf{\dimc{}}: \textit{why} does the developer select this particular API for their application's domain and does the API's domain align with the application's domain?
  \item \dimcat{D}~\textbf{\dimd{}}: \textit{what} additional API documentation can the developer find to aid their productivity?
  \item \dimcat{E}~\textbf{\dime{}}: is the \textit{visualisation} of the above information well organised and easy for the developer to digest?
\end{itemize}
Further descriptions of the categories encompassing each dimension are given within \cref{tse2020:fig:taxonomy,tse2020:tab:taxonomy}, coded as [$Xi$], where $i$ is the category identifier within a dimension, $X$, where $X~\in~\{ A, B, C, D, E \}$.

\Cref{tse2020:tab:taxonomy} shows which of the primary sources (S1--21) reports aspects of the artefacts described as an `in-literature score' (ILS). This score is calculated as a percentage of the number of primary studies that investigated or reported various issues regarding the specific artefact divided by the total of primary studies (see \cref{tse2020:sec:tax-analysis:ils}). This score is contrasted to the `in-practice score' (IPS) which indicates the overall level of agreement that \textit{practitioners} think such documentation artefacts are needed (see \cref{tse2020:sec:tax-analysis:ips}). 
For comparative purposes, we illustrate a colour scale (from red to green) to indicate the relevancy weight between ILS and IPS values in \cref{tse2020:tab:taxonomy} as per their assigned, discretised intervals (see \cref{tse2020:tab:ils-ips-intervals}). We also show illustrative interpretations of these generalised artefacts through italicised examples within \cref{tse2020:tab:taxonomy}.
We then provide three columns that assesses the presence of these documentation artefacts against three popular computer vision services: Google Cloud Vision, AWS's Rekognition, and Azure Cloud Vision (abbreviated to GCV, AWS and ACV). A fully shaded circle~(\circlepresent{}) indicates that the documentation artefact was clearly found in the service, while a half-shaded circle~(\circlepartialpresent{}) indicates that the artefact was only partially present. An outlined circle~(\circlenotpresent{}) indicates that the service lacks the indicated documentation artefact within our taxonomy. This empirical assessment is further detailed in \cref{tse2020:sec:tax-analysis:cvs-improvement}, which outlines concrete areas in the respective services' documentation where improvements could be made, as well as hyperlinks to the documentation where relevant. 

\Cref{tse2020:fig:taxonomy} illustrates a condensed version of taxonomy.
We provide iconography for the presence~(\faCheckCircle) or non-presence~(\faTimesCircle) of these   artefacts in \textit{all three} computer vision services assessed, per \cref{tse2020:sec:tax-analysis:ips}.

\section{Validating the Taxonomy}
\label{tse2020:sec:validation}

\subsection{Survey Study}
\label{tse2020:sec:validation:survey}

\subsubsection{Designing the Survey}

We followed the guidelines by \citet{Kitchenham:2007ux} on conducting personal opinion surveys in software engineering to validate our survey. In developing our survey instrument, we  shaped questions around each of our 5 dimensions and 34 categories. To achieve this, we used \citeauthor{Brooke:1996ua}'s SUS \citep{Brooke:1996ua} as a loose inspiration and re-shaped the 34 categories around a question that imitates the style of wording of questions used in the SUS. Each dimension was marked a numeric question (Q\#3--7), and alphabetic sub-questions were marked for each sub-dimension or category.

We used closed questioning where respondents could choose an answer on a 5-point Likert-scale (1=\textit{strongly disagree}, 2=\textit{somewhat disagree}, 3=\textit{neither agree nor disagree}, 4=\textit{slightly agree} and 5=\textit{strongly agree}).  Like \citeauthor{Brooke:1996ua}'s study, each question alternated in positive and negative sentiment. Half of our questions were written where a likely common response would be in strong agreement and vice-versa for the other half, such that participants would have to ``read each statement and make an effort to think whether they would agree or disagree with it'' \citep{Brooke:1996ua}. For example, the question regarding \dimcat{B7} on API limitations was framed as: ``\textit{I believe it is important to know about what the limitations are on what the API can and cannot provide}'' (Q4g), whereas the question regarding \dimcat{C1} on domain concepts of the API was framed as: ``\textit{I wouldn't read through theory about the API's domain that relates theoretical concepts to API components and how both work together}'' (Q5a). 
 
In addition, the remaining eight questions asked demographical information. An extra open question asked for further comments. The full survey is provided in \cref{tse2020:sec:survey} and anonymised survey data is available at \mbox{\url{https://bit.ly/33siqll}}.

\begin{figure*}[t]
\centering
\includegraphics[width=.9\linewidth]{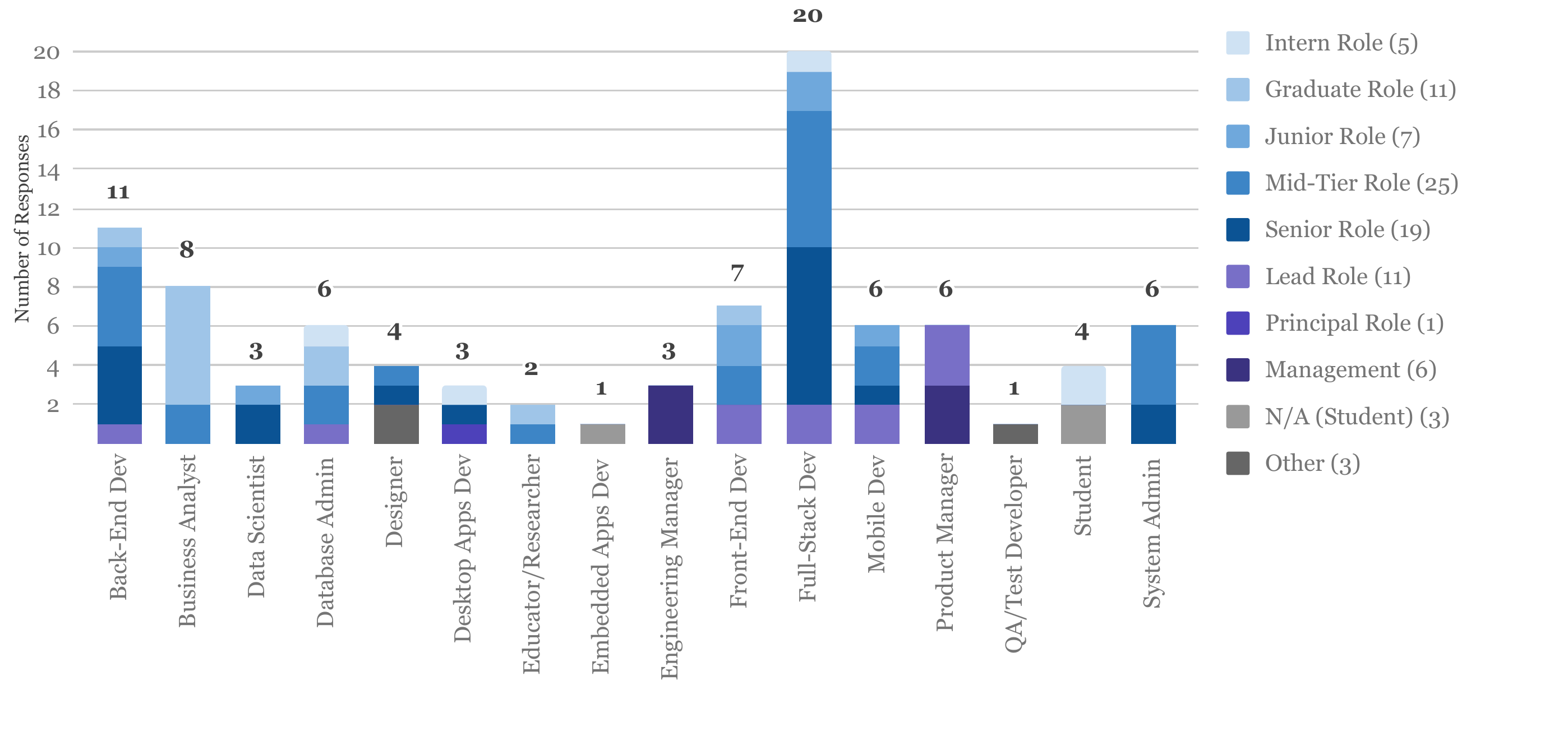}
\caption[Roles and seniority from survey participants]{A wide variety of roles and seniority were observed in our respondents.}
\label{tse2020:fig:roles-and-seniority}
\end{figure*}

\subsubsection{Evaluating the Survey}
\label{tse2020:sec:validation:survey:eval}

After the first pass at designing questions was completed, we evaluated our survey on three researchers within our research group for general feedback. This resulted in minor changes, such as slight re-wording of questions and providing specific questions with examples (some with images). For example, the question regarding \dimcat{A9} on an exhaustive list of all major components in the API was framed as ``\textit{I believe an exhaustive list of all major components in the API without excessive detail would be useful when learning an API}'' (Q3i) with the example ``\textit{e.g., a computer vision web API might list object detection, object localisation, facial recognition, and facial comparison as its 4 components}''.

After this, we conducted reliability analysis using a test-retest approach on three developers within our group seven weeks apart. Using the R statistical computation environment \citep{RCoreTeam}, we conducted our analysis using the \texttt{irr} package~\citep{Gamer:tj} (as suggested in~\citep{Hallgren:2012kt}) and  resulted in an average intra-class correlation (ICC) of 0.63 which indicates a good overall index of agreement \citep{cicchetti1994guidelines}.

\subsubsection{Recruiting Participants}

Our target population for the study was application software developers with varying degrees of experience (including those who and who have not used computer vision services or related tools before) and varying understanding of fundamental machine learning concepts. We began by recruiting software developers within our research group using a group-wide message sent on our internal messaging system. Of the 44 developers in our group's engineering cohort,\footnote{Our research group's engineering cohort consists of fully-qualified software engineers, with on average 5+ years industry experience.} \SurveyParticipantsInternal{} responses were returned, indicating an internal response rate of \SurveyParticipantsInternalResponseRate{}. Based on the \SurveyParticipantsInternal{} results from this internal trial, we calculated the median time to our complete survey was just over 20 minutes.

For external participant recruiting, we shared the survey on social media platforms and online-discussion forums relevant to software development. We adopted a non-probabilistic snowballing sampling where the participants, at the end of the survey, were encouraged to share the survey link to others using \textit{AddThis}.\footnoteurl{https://www.addthis.com/}{7 January 2020} Additionally, snowballing sampling was encouraged within members of our research group who were asked to share the survey. This sampling approach resulted in \SurveyParticipantsExternalSnowball{} external responses. A further \SurveyParticipantsExternalMTurk{} participants were recruited via Amazon Mechanical Turk\footnoteurl{https://www.mturk.com/}{9 July 2020}---often referred to as MTurk---which has been a successful approach adopted in previous software engineering surveys (e.g., \citep{Jiarpakdee2020}). To ensure our target demographic was selected, we applied the participant filter option `Employment Industry - Software \& IT Services'.
An additional \SurveyParticipantsExternalPartialResponses{} responses were partially filled (on average at a completion rate of \SurveyParticipantsExternalPartialResponsesCompletionRate{}). These partially completed responses were included in our analysis since they did yield some insight (see \cref{tse2020:sec:limitations:internal}). 
As participants recruited via MTurk have a financial incentive to complete surveys,\footnote{A total budget of AUD\$600 was allocated for recruitment via MTurk, with each participant receiving between AUD\$3.50--\$10.00.} we ensured strict quality control was applied to each survey response we received. For example, \SurveyParticipantsExternalResponseTooSmall{} participants opened the survey but did not answer any questions; for this reason, all survey responses by these participants were discarded. We identified that \SurveyParticipantsExternalMTurkRejected{} MTurk responses were filled out too quickly (where the median response time was under five minutes; well below the internal average of 20 minutes), and further analysis of these \SurveyParticipantsExternalMTurkRejected{} responses indicated poor reading of the question, and thus poor responses; this was identified via our use of alternating positively- and negatively-worded questions. Thus, \SurveyParticipantsExternalMTurkRejected{} MTurk responses were removed from the final analysis. Therefore, our final response rate yielded \SurveyParticipantsTotal{} responses of the total \SurveyParticipantsTotalReached{} participants reached; an overall response rate of \SurveyParticipantsTotalResponseRate{}.

\subsubsection{Analysing Response Data}
\label{tse2020:sec:validation:survey:analysis}

To analyse our response data, we produced a single score for each question's 5-point response. In line with with \citeauthor{Brooke:1996ua}'s SUS methodology \citep{Brooke:1996ua}, we subtracted one from the raw value of positive items, and subtracted the raw value from five for the negative items. This resulted in values on an ordinal scale of 0--4. We then averaged each response for every question and divide by four (i.e, now a 4-point scale) to obtain scores for each category. For example, two responses of \textit{strongly agree}=5 and one of \textit{neither agree nor disagree}=3 were given to \dimcat{A1} (positively worded); these values are mapped to 4 and 2, respectively, and are averaged (to 3.33) which is then divided by a maximum possible score of four, giving 0.84. We then discretise these calculated values into five intervals (as per \cref{tse2020:tab:ils-ips-intervals}, see \cref{tse2020:sec:tax-analysis:ips}) to interpret the findings; this is presented in \cref{tse2020:tab:taxonomy} under the `in-practice score' (IPS) for each category.

Demographics for our survey were consistent in terms of the experience levels of developers who responded. 78\% of respondents indicated they were professional programmers. Years of programming experience were: \textless 1~year (3.30\%); 1--5~years (41.76\%); 6--10~years (35.16\%); 11--15~years (9.89\%); 16--20~years (5.49\%); 21--30~years (3.30\%); 31--40~years (1.10\%); 41+ years~(0.00\%). A wide range of roles and seniority were listed by developers as presented in \cref{tse2020:fig:roles-and-seniority}, thereby indicating that our results include the different expectations of API documentation from a variety of sources. The highest role was a full-stack developer at either a mid-tier or senior role, followed by mid-tier or senior back-end developers and graduate and junior business analysts. Various managerial roles were also listed. Only five students (5.00\%) responded in our study, two listing themselves as interns with one as an embedded applications developer. Most respondents were Australian (40.00\%), Indian (26.70\%) or from the United States (20.00\%). Besides information technology services (30.77\%), consulting and other software development (both at 9.89\%) were the most predominant industries listed by participants.

\subsection{Empirical Application on Computer Vision Services}
\label{tse2020:sec:validation:empirical-app}

Once our taxonomy had been developed and assessed with developers, we performed an empirical application against three computer vision services: Google Cloud Vision \citepweb{GoogleCloud:Home}, Amazon Rekognition \citepweb{AWS:Home} and Azure Computer Vision \citepweb{Azure:Home}. Our selection criteria in choosing these particular services to analyse is based on the prominence of the service providers in industry and the ubiquity of their cloud platforms (Google Cloud, Amazon Web Services, and Microsoft Azure) in addition to being the top three adopted vendors used for cloud-based enterprise applications \citep{RightScaleInc:2018kJ}. In addition, we had conducted extensive investigation into the services' non-deterministic runtime behaviour and evolution profile in prior work \citep{Cummaudo:2019icsme} and have also identified developers' complaints about their incomplete documentation in a prior mining study on Stack Overflow \citep{Cummaudo:2020icse}.

We began with an exploratory analysis of the presence of each dimension and its categories. \Cref{tse2020:tab:docsources} displays all sources of documentation used; although we initially started on the respective services homepages \citepweb{GoogleCloud:Home,AWS:Home,Azure:Home}, this search was expanded to other webpages hyperlinked. For each category, we listed the documentation's presence as either fully present, partially present or not present at all. This is shown in \cref{tse2020:tab:taxonomy} with the indication of \mbox{(half-)filled} circles or circle outlines for Google Cloud Vision (abbreviated to GCV), Amazon Rekognition (abbreviated to AWS), and Azure Computer Vision (abbreviated to ACV). Notes were taken for each webpage justifying the presence, and exact sources of documentation were listed when (partially) present. PDFs of each webpage were downloaded between 14--18 March 2019 for analysis. Analysis was performed manually by the lead author by manual inspection of the downloaded web pages (as PDFs) and presence of each item was noted by the lead author using an approach similar to \citet{Watson:2012uy}.

\section{Taxonomy Analysis}
\label{tse2020:sec:tax-analysis}

In this section, we analyse investigating the taxonomy from two perspectives. Firstly, we contrast the ILS values, being an interpretation of the relevancy researchers have emphasised, against the IPS values found from the results of our survey (being an interpretation of what documentation artefacts developers value more). We are therefore able to identify the API documentation artefacts that are of high value to practitioners, but are yet to be deeply explored by researchers. Secondly, we contrast the IPS values against our assessment of computer vision services, and whether important API documentation artefacts have been included in popular services. We are therefore able to identify whether vendors have or have not already included these highly-valued documentation artefacts within their own APIs, and where existing areas of improvement lie.

\subsection{Exploring IPS and ILS Values}
\label{tse2020:sec:tax-analysis:ils-vs-ips}

\subsubsection{IPS Results}
\label{tse2020:sec:tax-analysis:ips}

\begin{table*}
  \centering
  \caption[Intervals assigned to ILS and IPS values]{Intervals of ILS (top) and IPS (bottom) values and frequencies.}
  \label{tse2020:tab:ils-ips-intervals}
  \begin{tabular}{c|ccl}
    \toprule
    \textbf{Research Attention} & \textbf{Range} & \textbf{Frequency} & \textbf{Categories}\\
    \midrule
Very Low & $0.00 \leq \textrm{ILS}(\,[Xi]\,) < 0.14$ & 7 & B4, B5, D6, B3, C1, D1, D2\\
Low & $0.14 \leq \textrm{ILS}(\,[Xi]\,) < 0.29$ & 13 & A1, A9, C3, D3, D4, E2, E3, E4, E5, B6, A7, A10, D5\\
Medium & $0.29 \leq \textrm{ILS}(\,[Xi]\,) < 0.43$ & 9 & B2, B7, A4, A12, E1, A3, A8, A11, C2\\
High & $0.43 \leq \textrm{ILS}(\,[Xi]\,) < 0.57$ & 3 & E6, B1, A2\\
Very High & $0.57 \leq \textrm{ILS}(\,[Xi]\,) \leq 0.71$ & 2 & A6, A5\\
    \midrule
    \midrule
    \textbf{Value to Developers} & \textbf{Range} & \textbf{Frequency} & \textbf{Categories}\\
    \midrule
Very Low & $0.00 \leq \textrm{IPS}(\,[Xi]\,) < 0.18$ & 0 & --\\
Low & $0.18 \leq \textrm{IPS}(\,[Xi]\,) < 0.36$ & 0 & -- \\
Medium & $0.36 \leq \textrm{IPS}(\,[Xi]\,) < 0.53$ & 6 & D4, B4, C3, C1, E4, B3\\
High & $0.53 \leq \textrm{IPS}(\,[Xi]\,) < 0.71$ & 16 & A4, B6, A2, D2, A6, E2, B5, D6, A8, B2, E6, A10, E5, D5, A9, D3\\
Very High & $0.71 \leq \textrm{IPS}(\,[Xi]\,) \leq 0.89$ & 12 & E3, A7, A3, C2, A12, B1, D1, A11, A1, E1, A5, B7\\
    \bottomrule
  \end{tabular}
\end{table*}

IPS values indicate the extent to which developers agree with the statements made in our survey, as calculated by the method described in \cref{tse2020:sec:validation:survey:analysis}. The interpretation of these values are the documentation artefacts (categories) that developers \textit{value} the most. Thus collectively, these artefacts indicate the overall level of importance towards specific API documentation requirements (dimensions). 

To interpret these values, we group the data from each of our survey's 34 statements (for each category) into an ordinal scale of five intervals. These intervals indicate relative value to developers; a documentation artefact has \textit{very low} value to developers, \textit{low} value, \textit{medium} value, \textit{high} value, or \textit{very high} value. \Cref{tse2020:tab:ils-ips-intervals} presents these intervals and frequencies of each, with the order of the categories shown in the last column indicating raw IPS values (least useful to most useful) before discretisation in ascending order.

Practitioners tend to agree that each documentation artefact is important to have, and thus IPS values likely fall into the \textit{High} or \textit{Very High} intervals. Only six categories fall into the \textit{Medium} interval and none fall into lower intervals. Developers find technical support contact information \dimcat{D4} to be of the lowest value (see \cref{tse2020:tab:ils-ips-intervals}), likely since developers tend to rely on crowd-sourced peer support through mediums such as Stack Overflow. They also see little value in: descriptions of the types of end-users the API is intended for \dimcat{B4}; documentation for non-technical audiences \dimcat{C3}; conceptual information relating the API back to its application domain \dimcat{C1}; structured navigation of the presented API documentation \dimcat{E4}; and descriptions of the intended developers who should be using the API \dimcat{B3}.

\subsubsection{ILS Results}
\label{tse2020:sec:tax-analysis:ils}
ILS values indicate overall research attention of categories of our taxonomy through the proportion of papers in our SMS that investigated or reported various issues regarding a specific API documentation artefact. Collectively, each of these categories combined form a dimension (labelled A--E) in a bottom-up approach (see \cref{tse2020:sec:method:taxonomy-development:design-phase}). Each dimension (top-node) describes the requirements of good quality API documentation, while the category (leaf-node) is the specific API documentation artefact that, collectively, form the requirement. A category with a high ILS value indicates that existing studies that there is substantial attention by researchers on this specific documentation artefact (or, collectively, requirement of good quality API documentation). Conversely, a lower ILS value indicates less attention reported on these categories (artefact) or dimensions (requirement) by the software engineering research community.

To demonstrate the attention of these documentation artefacts within literature, we interpret the ILS values in a similar fashion to the IPS values. It is represented as a discretised value of intervals within a five-dimensional ordinal scale, where the attention on these artefacts in literature are one of: \textit{very low} attention, \textit{low} attention, \textit{medium} attention, \textit{high} attention, \textit{very high} attention. \Cref{tse2020:tab:ils-ips-intervals} indicates the boundaries for each interval (as calculated by the highest ILS value of \dimcatILSvalueAFive{} divided by the five intervals) in addition to the frequency of categories appearing in each interval. The order of the categories shown in the last column indicate the ascending order (least research attention to most) of raw ILS values before discretisation. As shown, most of the artefacts (20) found in the taxonomy are discussed in literature disproportionately more than others (i.e., those that fall into the `low' (13) or `very low' (7) intervals), though the underlying reasons behind this should be considered on a case-by-case basis (see \cref{tse2020:sec:threats:construct}.

There are only five categories that fall into the `high' or `very high' intervals, three of which fall under dimension \dimcat{A}, \dima{}. Research attention on a particular documentation artefact that is considered \textit{Very High} gravitates towards code snippets \dimcat{A5} and tutorials \dimcat{A6}. Code snippets are the readiest form of API documentation for developers, representing exemplary nuggets of information for developers to rapidly digest singular components of the API's functionality. While code snippets generally only reflect small portions of API functionality (generally limited to 15--30 LoC), this is complimented by step-by-step tutorials. These may tie in multiple (disparate) components of API functionality to demonstrate development of more non-trivial applications. Therefore, unsurprisingly, research has substantially explored how best API developers can extract code snippets or write tutorials for these purposes in mind. This is followed by low-level reference documentation \dimcat{A2}---under the `high' interval---whereby developers should document all client-facing implementation or usage aspects of their API (e.g., class, method, parameter descriptions etc.).
Lastly, the entry-level purpose/overview of an API \dimcat{B1} and consistency in the look and feel of the documentation throughout all of the API's official documentation \dimcat{E6} are fall under the `high' interval. API vendors must give motivation as to why a developer should choose a particular API over another, articulating the \textit{need} of their API, presenting this and other documentation aspects in the easiest way for developers to consume.

\subsubsection{Research Opportunities for High-Value Artefacts}
\label{tse2020:sec:tax-analysis:ips-vs-ils}

\begin{figure}[h]
  \includegraphics[width=\linewidth]{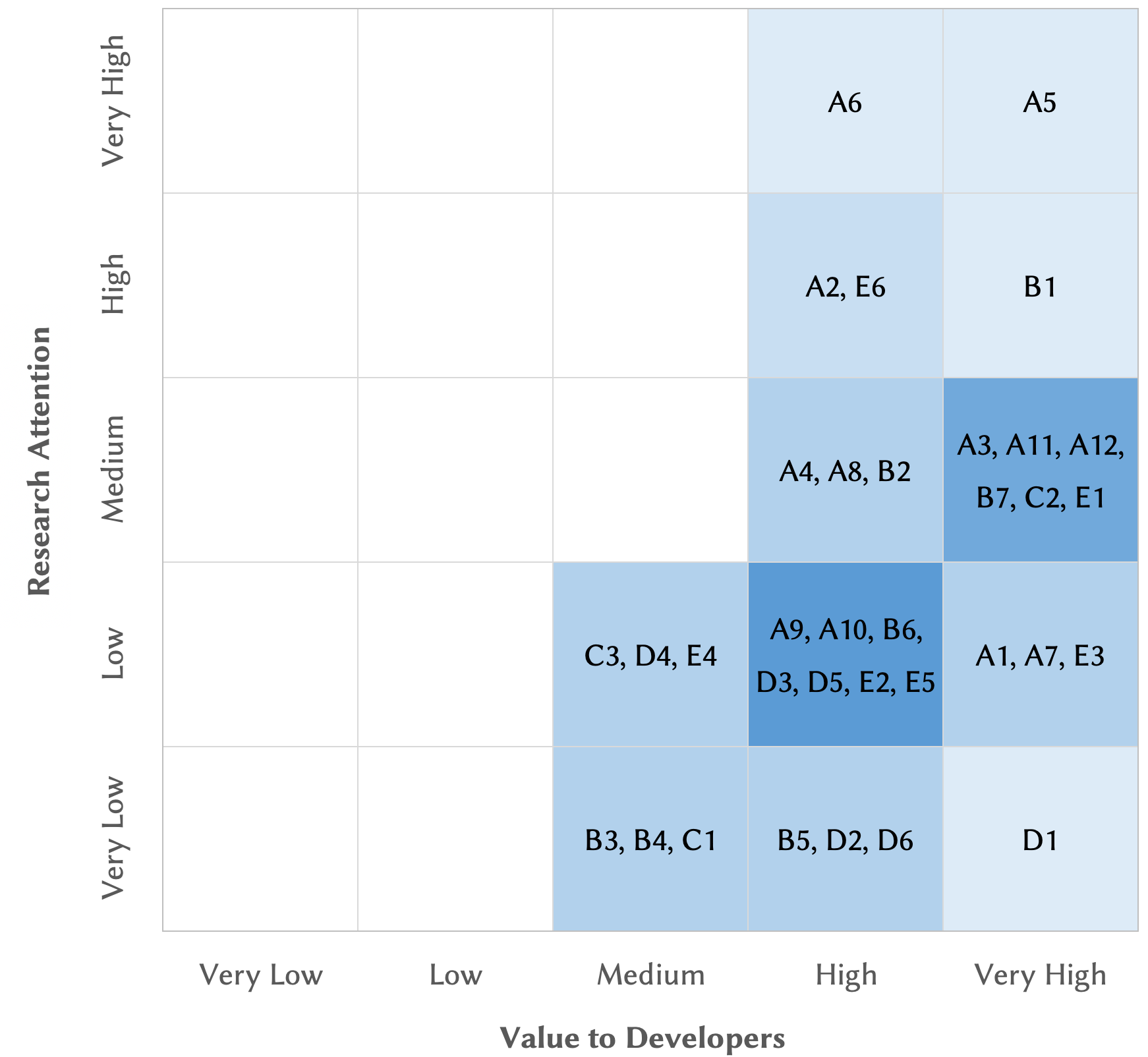}
  \caption[Comparing value of API documentation artefacts to developers vs research attention]{Value of API documentation artefacts to developers (IPS) vs their research attention (ILS). Colour intensity represents greater number of categories in each intersection}
  \label{tse2020:fig:ips-vs-ils}
\end{figure}

In this section, we explore the ILS and IPS values as two distinct indicators of research exploration that would provide the most value to practitioners. We then provide a qualitative discussion by inspecting the intersection of categories at each respective interval identified by our SMS and survey study. Thus, we are able to determine documentation artefacts (categories) and requirements (dimensions) that provide the \textit{greatest value} to developers but have not gained proportional attention in the software engineering literature when compared to other artefacts, and vice-versa. Graphically, we represent these intersections within a five-by-five matrix with intervals of the IPS ($x$ axis) plotted against intervals of the ILS ($y$ axis). Intersections between the two are listed for each category within the taxonomy. This is presented in \cref{tse2020:fig:ips-vs-ils}.

There is a distinction between \mbox{(very-)highly} valued documentation artefacts whose research attention is \mbox{(very-)low}, as presented in the bottom-right of \cref{tse2020:fig:ips-vs-ils}. Most notably, we find that developers find \dimd{} \dimcat{D} a highly valued API documentation requirement, but there still exists a substantial gap in existing literature into this requirement. For example, besides category \dimcat{D4} (which is of only \textit{Medium} value to developers), less research has explored all other dimension \dimcat{D} categories (though there may be understandable reasons as to why, as detailed in \cref{tse2020:sec:threats:construct}). Furthermore, developers highly value detailed \dima{} \dimcat{A} through many documentation artefacts, notably quick-start guides \dimcat{A1}, downloadable sample applications \dimcat{A7}, exhaustive list of major components \dimcat{A9}, and system requirements to use the API \dimcat{A10}. Such artefacts emphasise the need for developers to rapidly pick-up a new API; however, the best ways to provide such information is still open to further investigation in literature.

Conversely, the top-right of \cref{tse2020:fig:ips-vs-ils} emphasises (very)-highly researched artefacts that are of (very)-high value to developers. Here we see that \dima{} \dimcat{A} is the most-researched requirement, with code snippets \dimcat{A5} being an API usage artefact that is both most-researched and of highest value. Hence, this demonstrates how many existing studies have an empirical basis on software developers (e.g., via surveys or interviews; see \cref{tse2020:fig:sms})---code snippets is a well-researched artefact since most developers agree to its need in the documentation of APIs. Therefore, it is clear to see how the correlation between the respective ILS and IPS values for \dimcat{A5} are high. However, if we look at other areas of our taxonomy, such as \dimcat{A12},  \dimcat{B7}, \dimcat{D3}, \dimcat{E3} or \dimcat{E5}, we find that developers do indeed desire these aspects of API documentation, and, consequently, demand usage descriptions, design rationale descriptions, support artefacts, or good presentation of the documentation to be a necessary requirement of good quality API documentation. Thus, these aspects have not gained proportional attention in literature, thereby highlighting future research potential.

\subsection{Triangulating IPS, ILS and Computer Vision}

\begin{figure}[h]
  \includegraphics[width=\linewidth]{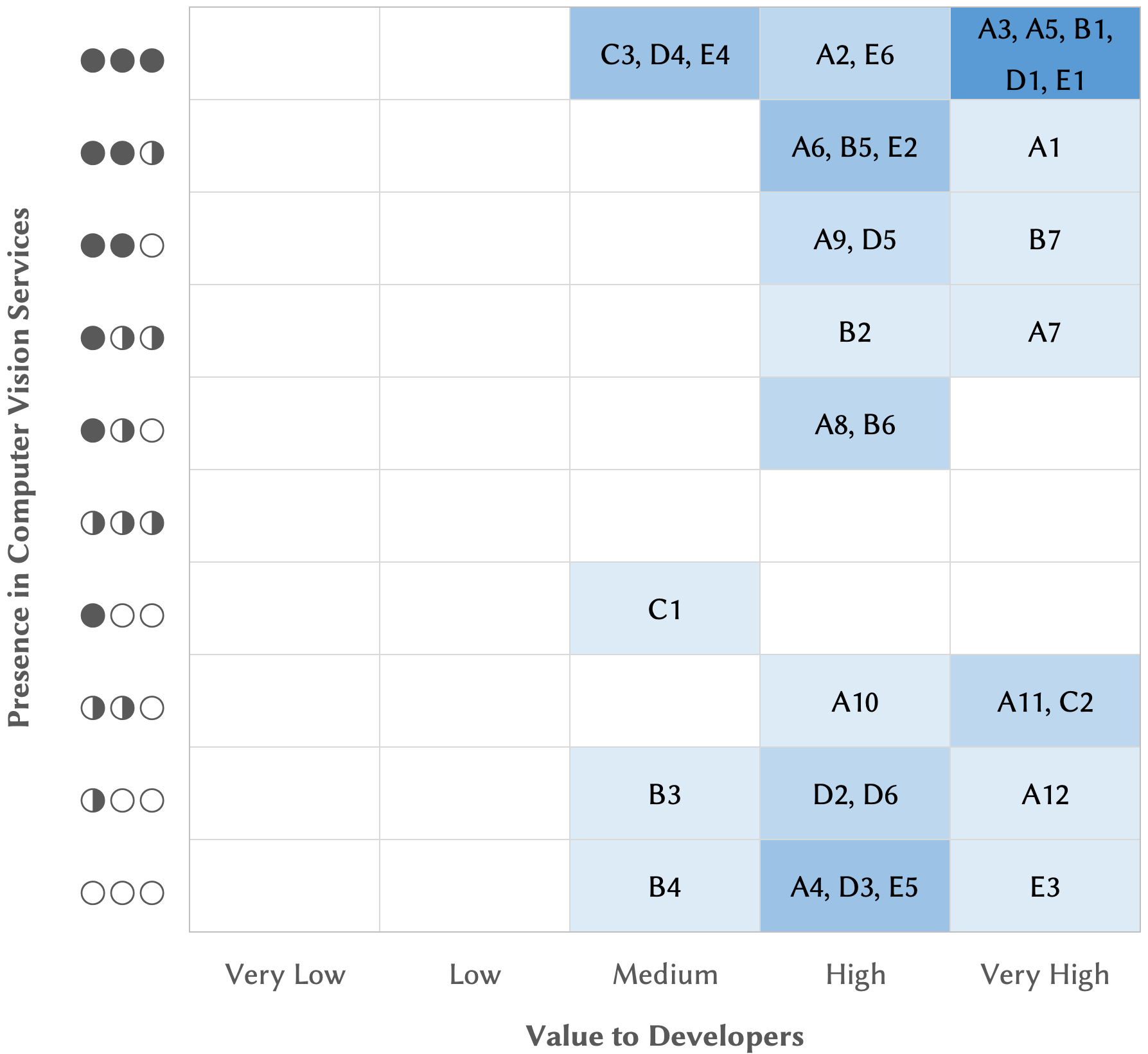} 
  \caption[Comparing value of API documentation artefacts to presence in computer vision services]{Value of API documentation artefacts to developers (IPS) vs their presence in computer vision services. Colour intensity represents greater number of categories in each intersection.}
  \label{tse2020:fig:ips-vs-cvs}
\end{figure}

To interpret our comparison of IPS values with computer vision services, we introduce a calculated `presence score' for each category. As discussed in \cref{tse2020:sec:validation:empirical-app}, we empirically evaluate each category of our taxonomy with three computer vision services: Azure Computer Vision (ACV), Amazon Rekognition (AWS) and Google Cloud Vision (GCV). We indicate whether the respective API documentation artefact is present, partially present, or nor present (as listed in \cref{tse2020:tab:taxonomy}). To interpret this data, we assign a full circle (\circlepresent{}) for present, half-circle (\circlepartialpresent{}) for partially present and an empty circle (\circlenotpresent{}) for not present. Combinations of presence for each category per service are indicated with the three circles of varying shade. For example, \dimcat{A1} has a presence score of {\small \circlepresent{}~\circlepresent{}~\circlepartialpresent{}} because it was found to be present in both GCV and ACV but only partially present in AWS; \dimcat{B3} has a presence score of {\small \circlepartialpresent{}~\circlenotpresent{}~\circlenotpresent{}} because it was only found to be partially present in GCV, etc. For a list of full presence values, see \cref{tse2020:tab:taxonomy}.

We illustrate which artefacts industry vendors provide developers with and the artefact's respective developer value using this combination of three circles. Using a similar approach to the previous section, these results are presented in a ten-by-five matrix as illustrated in \cref{tse2020:fig:ips-vs-cvs}. If only one service fully implements a documentation artefact of \mbox{(very-)high} value to developers ({\small \circlepresent{}~\circlenotpresent{}~\circlenotpresent{}}), if one or two services partially implement the artefact ({\small \circlepartialpresent{}~\circlenotpresent{}~\circlenotpresent{}} and {\small \circlepartialpresent{}~\circlepartialpresent{}{}~\circlenotpresent{}{}}) or if none do ({\small \circlenotpresent{}~\circlenotpresent{}~\circlenotpresent{}}), then we believe there is room for improvement for service vendors to improve their documentation and include these artefacts. 

In this instance, we can see 10 categories listed in \cref{tse2020:fig:ips-vs-cvs} that developers feel are important but are not fully implemented across all three computer vision service vendors. This is especially the case for dimensions  \dimcat{A} (\dima{}) and \dimcat{D} (\dimd{}), corroborating our findings with existing gaps in literature under \cref{tse2020:sec:tax-analysis:ips-vs-ils}. In other words, while both the goals of existing studies and computer vision service vendors have emphasised the need for artefacts such as code-snippets \dimcat{A5}, tutorials \dimcat{A6}, and entry-points to the API \dimcat{B1}, less attention is given to by \textit{both} literature and vendors on the same, \mbox{(very-)highly} valued aspects to developers (e.g., troubleshooting hints \dimcat{D2}, licensing information \dimcat{D6} or links to related components \dimcat{E3}).

Furthermore, from our analysis, we can see areas with which the research community has and has \textit{not} paid extensive attention to. We still see that vendors have paid attention to artefacts even where there has been less research attention, namely \dimcat{D1} (FAQs), \dimcat{B5} (success stories), \dimcat{A7} (downloadable sample applications), \dimcat{A1} (quick-start guides), \dimcat{E2} (forums), \dimcat{D5} (printable guides), and \dimcat{A9} (API component lists). These seven categories are of (very) high value to developers but research attention on these topics are (very) low; however, their presence score within computer vision services are {\small\circlepresent{}{}{} \circlepartialpresent{}~\circlepartialpresent{}{}} or greater. Hence, we can see that vendors address developer's concerns despite the lack of attention by software engineering researchers in these areas, and thus future research potential to better serve developers and ensure vendors' implementation of these documentation artefacts is evident. 

From the above, we can therefore conclude that the vendors' documentation largely covers a majority of API documentation requirements. However, there still remains opportunity for improvement to API documentation by either vendors and/or the research community: that is, low research attention on documentation artefacts that present high value to developers which are \textit{also} generally missing from vendor documentation. To explore this aspect, we triangulate the documentation artefacts (categories) that have a low or very low research attention and that are only present in one service, partially present in one or two, or not present at all. This results in three documentation requirements that warrant further exploration by industry vendors or the research community (see \cref{tse2020:tab:high-ips-low-ils-low-cvs}).

\begin{table*}
  \centering
  \caption[Documentation artefacts of high value to developers that are under-researched and under-documented]{Documentation artefacts of high value to developers that have less attention in software engineering literature and are under-documented in computer vision services. Documentation requirements (i.e., dimensions) separated by rules.}
  \label{tse2020:tab:high-ips-low-ils-low-cvs}
  \begin{tabular}{p{0.325\linewidth}|lp{0.15\linewidth}p{0.35\linewidth}}
    \toprule
    \textbf{Artefact} & \textbf{Value} & \textbf{Research Attention} & \textbf{Presence in Computer Vision Services}\\
    \midrule
    \dimcat{A10} Documenting API's minimum system requirements and/or dependencies & 
    High &
    \textbf{Low:} 5 studies (23\%) &
    \textbf{Score=1.0:} No dedicated web pages found for this artefact in any service. Dependencies for client libraries embedded within GCV and ACV quick-start guides \citepweb{Quicksta92:online,CalltheC0:online}. Other system requirements not listed.\\
    \midrule
    \dimcat{D2} Troubleshooting hints & 
    High &
    \textbf{Very Low:} 2 studies (10\%) &
    \textbf{Score=0.5:} Only found in AWS's video recognition service \citepweb{Troubles2:online}, but no troubleshooting tips found for non-video image recognition.\\
    \dimcat{D3} Diagrammatic representation of API & 
    Very High &
    \textbf{Low:} 3 studies (14\%) &
    \textbf{Score=0.0:} Not found for any service.\\
    \dimcat{D6} Licensing Information & 
    Very High &
    \textbf{Very Low:} 1 study (5\%) &
    \textbf{Score=0.5:} Partially present only in ACV \citepweb{LegalTermsMS:online}; information is non-specific to the licensing terms of ACV exclusively.\\
    \midrule
    \dimcat{E3} Quick-links to other relevant components & 
    Very High &
    \textbf{Low:} 3 studies (14\%) &
    \textbf{Score=0:} Not found for any service.\\
    \dimcat{E5} Visualised map of navigational paths & 
    Very High &
    \textbf{Low:} 3 studies (14\%) &
    \textbf{Score=0:} Not found for any service.\\
    \bottomrule
  \end{tabular}
\end{table*}

\subsection{Recommendations Resulting from Analysis}
\label{tse2020:sec:tax-analysis:cvs-improvement}

In this section, we triangulate the taxonomy developed from literary sources, the developer survey on this taxonomy to understand its efficacy in-practice, and the application of the taxonomy to computer vision services to provide several recommendations for both service providers and researchers. Our recommendations are based both on extrapolations of our findings, our prior work, and existing experience with such work. 

\subsubsection{Recommendations for vendors}

\Cref{tse2020:tab:high-ips-low-ils-low-cvs} emphasises how service vendors still lack key documentation requirements of critical importance to developers that are still widely under-researched in software engineering literature. The largest of these requirements are the need for vendors to provide additional support artefacts \dimcat{D} and the need for vendors to present this in a way that's most digestible for developers to understand \dimcat{E}. A list of detailed suggestions for vendors are provided in \cref{tse2020:sec:suggested-improvements}; here we discuss generalised findings on a sample of key artefacts.

For example, no services assessed had any form of diagrammatic overview of their APIs at a high-level \dimcat{D3}, thereby indicating how various components of their APIs work together, such as how specific endpoints work or an overview of the lifecyle of the technical domain behind these endpoints (i.e., label/train/infer/re-train), thereby incorporating conceptual relationships behind the API \dimcat{C1}. For instance, an interactive overview of the developer's need to pre-process their data, send it to the service, and post-process the response data would help developers understand how the service better fits into the `flow' of their application. Moreover, we failed to find lower-level diagrammatic overviews of the client SDKs---such as a UML diagram---that developers find very useful. We strongly advise vendors to provide diagrams illustrating the service within context to help support existing written documentation.

Troubleshooting hints \dimcat{D2} are also a valuable support artefact, but were only found for AWS's video processing endpoints. As our prior work shows, developers are likely to question what aspects of the service can and cannot do, such as the types of labels it can find, or how to make it focus on specific ontologies when an input image is provided; e.g., time of day (day vs night) location (indoors vs outdoors) or the subject of the image (dog vs cat) \citep{Cummaudo:2020icse}. Troubleshooting in identifying service evolution \citep{Cummaudo:2020icse} would also be important, since developers are likely to overlook subtle (but application-breaking) changes to response data, such as labels introduced/removed or confidence changes. Therefore, vendors must document detailed troubleshooting suggestions on their websites on how best to resolve discrepancies in the results found from these services. This could easily be tied in with \dimcat{A12} to incorporate usage description requirements when errors are presented to users and how to deal with them; also largely missing from existing documentation.

Another important aspect is the need to make documentation of one component more easily relatable to other parts of the documentation \dimcat{E3}. Again, no service provided quick-links to related documentation; an example here could be links to definitions of domain-specific terminology \dimcat{C2} to help developers with the learning process of adopting these new generation of APIs (e.g., the `score' field could be linked back to a video explaining the concept of probability within the services' guesses).

\subsubsection{Recommendations for researchers}
As shown in \cref{tse2020:tab:high-ips-low-ils-low-cvs}, we see that there are cases of (very) high-value documentation artefacts (to practitioners) in which literature has not paid great attention to. For example, for the requirement of API usage description \dimcat{A}, practitioners agree that both code snippets \dimcat{A5} and documenting system requirements to use the API \dimcat{A10} are of, at least, high value. However, while code snippets has had \textit{consistent} attention within the software engineering research community (i.e., 15 papers spanning 1998--2019), we see that system requirements documentation only gained fluctuating interest by researchers (i.e., predominantly in the 2000s, with two further papers in the last three years). Thus, five papers investigating \textit{some} aspects on this artefact may not cover \textit{all} its aspects; for example, we may have identified a \textit{need} to document these requirements and dependencies, but does this mean we know \textit{all} aspects on how to produce them, the best way to \textit{communicate} them, and the most efficient means for developers to \textit{consume} that information? Contrasting this artefact against the 15 papers on code snippets, we see two documentation artefacts of at least high value to practitioners, yet, evidently, researchers have paid attention to one over the other. 

As \cref{tse2020:fig:ips-vs-ils} shows, the need for additional support \dimcat{D} within documentation is the largest requirement that \textit{may} be an indicator for further research in this domain (see \cref{tse2020:sec:threats:construct}). Notably, RQ2 of our SMS identified the methodologies and data collection techniques by which our existing understanding of API documentation requirements were gathered; as demonstrated through \cref{tse2020:fig:sms}, a majority of our understanding is grounded through the opinions of developers, namely evaluation research using direct techniques. Too many studies are shown to rely on a handful of data collection techniques (interviews and questionnaires, shadowing and observation, think-aloud sessions) and a stronger emphasis for indirect and independent techniques is needed moving forward; there is therefore a gap in literature on \textit{other} types of data collection techniques that may provide different insights into satisfying the documentation requirements within our taxonomy.

For example, we see \dimcat{A9} (exhaustive list of major API components) as a high-value documentation artefact that satisfies the requirement of the API usage description \dimcat{A}. However research attention is lower. A validation research paper could propose a method to generate a baseline list of these components through an independent technique, such mining the API codebase for its major components through class usage (static analysis) or analysing an existing work database or tool use logs to see which components developers have accessed the most. This would satisfy the need for the documentation artefact, bolstering the API usage requirement and exploring new techniques to do so.

Few philosophical papers result in a lack of insight into completely new ways of exploring API documentation. Further exploration into this type of research may help us devise a whole new framework of producing API documentation. For example, as shown by developers and vendors, quick-start guides \dimcat{A1} are highly valued, and well-documented in computer vision services. But literature does not provide any vocabulary or frameworks into how best to develop such guides. Involving both software engineering researchers and developers through a brainstorming or focus group to conceptualise, devise, and refine such a framework may be a worthwhile study to better improve our understanding of quick-start guides whilst also exploring new approaches to research new guidelines.

Beyond requirement \dimcat{A}, another insight identified is the need for developers to have visualised maps of navigational paths \dimcat{E5} which is not yet provided by any of the computer vision service providers investigated. With the low ILS value in this category (14\% or 3 studies), we see a potential research topic for future exploration. For example, if research can demonstrate that such visualised maps are not just something developers desire, but can make them \textit{more effective} in their day-to-day work, then this could be a strong case made to vendors to improve the presentation of their documentation. 

Thus, as we have shown in these sample recommendations, many potential studies and research directions can stem by exploring the discrepancies of API documentation in literature, in practice, and their presence in computer vision services (i.e., as a sample case study) when assessed on a case-by-case basis. The method researchers decide upon depends the research questions they wish to address; thus, observations we present in \cref{tse2020:fig:sms} may trigger fruitful reasoning about approaches future research could take, however inferring methodological gaps will need to be compatible with research goals. Thus, mapping these discrepancies to gaps in the techniques used in studies to devise of novel ways to improve API documentation whilst also exploring new methodologies should be balanced carefully by researchers.

\section{Threats to Validity}
\label{tse2020:sec:limitations}

\subsection{Internal Validity}

Threats to \textit{internal validity} represent internal factors of our study which affect concluded results. \citeauthor{Kitchenham:2007dd}' guidelines on producing systematic reviews \citep{Kitchenham:2007dd} suggest that researchers conducting reviews should discuss the review protocol, inclusion decisions, data extraction with a third party. Within this study, we discussed our protocols with other researchers within our research group and utilised test-retest reliability. Further assessments into reliability would involve an assessment of the review and extraction processes, which can be investigated using inter-rater reliability measures. Guidelines suggested by \citet{Garousi:2017:EGE:3084226.3084238} describe methods for independent analysis and conflict resolution could help resolve this.

As stated in \cref{tse2020:sec:method:taxonomy-development}, we utilised a systematic SE taxonomy development method by \citet{Usman:2017hn}. Two additional taxonomy validation approaches proposed by \citeauthor{Usman:2017hn} were not considered in our work: benchmarking and orthogonality demonstration. To our knowledge, there are no other studies that classify existing API documentation studies into a structured taxonomy, and therefore we are unable to benchmark our taxonomy against others. We would encourage the research community to conduct a replication of our work and investigate whether our taxonomy classification approaches are replicable to ensure that categories are reliable and the dimensions fit the objectives of the taxonomy. Moreover, we did not investigate orthogonality demonstration as our primary goals for this work were to investigate the efficacy of the taxonomy by practitioners and in-practice, with reference to our wider research area of intelligent computer vision services. Therefore, we solely adopted the utility demonstration approach in two detailed experiments (\cref{tse2020:sec:validation,tse2020:sec:tax-analysis}) to analyse the efficacy of our taxonomy and identify potential improvements for these services' API documentation.

\subsection{External Validity}\label{tse2020:sec:limitations:internal}

Threats to \textit{external validity} concern the generalisation of our observations. Our systematic mapping study has used a broad range of sources however not all papers contributing to API documentation may have been found or captured within the taxonomy. While we attempted to include as many papers as we could find in our study, some papers may have been filtered out due to our exclusion criteria. For example, there are studies we found that were excluded as they were not written in English, and these excluding factors may alter our conclusions, introducing conflicting recommendations. However, given the consistency of these trends within the studies that were sourced, we consider this a low likelihood.

Online documentation of APIs are non-static, and may evolve using contributions from both official sources and the developer community (e.g., via GitHub). We downloaded the three service's API documentation in March of 2019---it is highly likely that new documentation may have been added since or modified since publication. A recommendation to mitigate this would be to re-evaluate this study once intelligent computer vision services have matured and become even more mainstream in developer communities.

Unless significant inducements are offered, \citet{Singer:2007tu} report that a consistent response rate of 5\% has been found in software engineering questionnaires distributed and in information systems the median response rates for surveys are 60\% \citep{Baruch:1999vf}. We observe that low response rates may adversely effect the findings of our survey, typically as software engineers find little time to do them \citep{Singer:2007tu}. When compared to typical software engineering studies, our response rate of \SurveyParticipantsTotalResponseRate{} was likely successful due to designing and carefully testing succinct, unambiguous and well-worded questions with  researchers within our research group. All adjustments made from the pilot study due to unexpected poor quality of the questionnaire have been reported and explained in \cref{tse2020:sec:validation:survey:eval}. However, further improvements could be made to increase this response rate.

The survey reached \SurveyParticipantsExternalTotal{} external and \SurveyParticipantsInternal{} internal participants. This yielded a total of \SurveyParticipantsTotal{} participants. However, only \SurveyParticipantsFullResponses{} participants fully completed the survey and, on average, those who only partially completed the survey completed \SurveyParticipantsExternalPartialResponsesCompletionRate{} of all questions. Therefore, demographic data for these participants is largely missing.
To verify the reliability of partially submitted responses, we calculated the average response of each item in our survey (i.e., question) for all fully completed results and all partially completed results. All partially completed questions, except \dimcat{B7}, were within 1 standard deviation from the mean, and therefore we believe the  \SurveyParticipantsExternalPartialResponses{} partial results to be valid when excluding B7. Even if these partial results are excluded, our full-response participant count of \SurveyParticipantsFullResponses{} is still comparable to existing studies, such as \citet{Nykaza:2002td} (57 participants), \citet{Robillard:2011uv} (80 participants), or \citep{Robillard:2009uk} (83 participants). Therefore, given these comparable numbers, we believe this does not compromise validity of our results.

We also adopt research conducted in the field of questionnaire design, such as ensuring all scales are worded with labels \citep{Krosnick:1999wt} and have used a summating rating scale \citep{Spector:1992uj} to address a specific topic of interest if people are to make mistakes in their response or answer in different ways at different times. This approach was also extended using alternating positive and negative sentiment for each question---as multiple studies have shown \citep{Sauro:2011aj,Brooke:2013vt}, this approach helps reduce poor-quality responses by minimising extreme responses and acquiescence biases. 

\subsection{Construct Validity}\label{tse2020:sec:threats:construct}

Threats to \textit{construct validity} relates to the degree by which the data extrapolated in this study sufficiently measures its intended goals. Our interpretation of the ILS (as given in \cref{tse2020:sec:findings,tse2020:sec:tax-analysis:ils}) is reported as the proportion of papers whose research investigates or explores issues regarding the aspects of specific API documentation artefacts (i.e., categories in the taxonomy) that, collectively, comprise the requirements of good API documentation (i.e., dimensions in the taxonomy). Every effort has been made in this work to provide a constructive analysis on the API documentation landscape, however, the studies that comprise the ILS may differ in their intent toward a specific documentation artefact. For example, some studies may have distinct goals to extensively study \textit{how} code snippets \dimcat{A5} specifically improve developer productivity (e.g., through interviews or by observational studies), while others may just reflect that code snippets are a commonly-used artefact self-reported by developers (e.g., through a survey). Thus, the interpretation of the ILS may range between deep exploration of an artefact or whether a study mentions the artefact without any attempts to thoroughly investigate it. For this reason, we suggest that a high ILS value for a category within the taxonomy suggests that the documentation artefact is within the attention of the research community, and that subsequent attention \textbf{may} be required for those artefacts with low ILS values as a \textit{potential indicator} for future research (i.e., it also may \textit{not}). However, each artefact with a low ILS (but high IPS) would need to be carefully examined in isolation to evaluate whether future research is indeed warranted, and how that research can be conducted with the ultimate goal to assist practitioners.

Automatic searching was conducted in the SMS by choice of three popular databases (see \cref{tse2020:sec:method:lit-review}). As a consequence of selecting multiple databases, duplicates were returned. This was mitigated by manually curating out all duplicate results from the set of studies returned. Additionally, we acknowledge that the lack manual searching of papers within particular venues may be an additional threat due to the misalignment of search query keywords to intended papers of inclusion. Thus, our conclusions are only applicable to the information we were able to extract and summarise, given the primary sources selected.

While we have investigated the application of this taxonomy using a user study (\cref{tse2020:sec:validation:survey}), we would like to explore a controlled study of developers to assess how improved and non-improved API documentation impacts developer productivity. The outcome of this work can help design a follow-up experiment, consisting of a comparative controlled study \citep{Seaman:2007wa} that capture firsthand behaviours and interactions toward how software engineers approach using a computer vision service with and without our taxonomy applied. This can be achieved by providing `mock' improved documentation with the suggested improvements included in this work. Such an experiment could recruit a sample of developers of varying experience (from beginner programmer to principal engineer) to complete a certain number of tasks under a comparative controlled study, half of which will (a) develop using the improved `mock' documentation, and the other half will (b) develop with the \textit{as-is/existing} documentation. From this, we can compare if the taxonomy makes improvements by capturing metrics and recording the sessions for qualitative analysis. Visual modelling can be adopted to analyse the qualitative data using matrices \citep{Dey:2003ty}, maps and networks \citep{Miles:1994ty} as these help illustrate any causal, temporal or contextual relationships that may exist to map out the developer's mindset and difference in approaching the two sets of designs of the same tasks.

\section{Conclusions \& Future Work}
\label{tse2020:sec:conclusions}

The emergence of AI-based intelligent components present significant challenges to our existing understanding of traditional API documentation. The inherent probabilistic and non-deterministic nature of these components means that developers must shift their mindset of conventional APIs, and vendors of these services must similarly shift the mindset of documenting their APIs using traditional means. Without adapting to the new mental model (of the vendors designing these services) and by vendors presenting poor or incomplete (traditional) documentation that is not compatible with these next-generation components, developers face many struggles. They fail to grasp how to properly understand how these services work, seeking further documentation or support from their peers on forums on such as Stack Overflow \citep{Cummaudo:2020icse}. This ultimately hinders developers' productivity and thus adversely affects the internal quality of the applications that they build.

This study has explored the artefacts and means by which traditional API documentation is studied through the use of an SMS of 4,501 studies, identifying 21 key works. From this, we synthesised a taxonomy of the various documentation artefacts that improves API documentation quality, and thus collectively synthesising the requirements of good API documentation. Furthermore, we also capture the most commonly used analysis techniques used in the academic literature to understand the means by which the goals of these studies resulted in their findings. We then validate our taxonomy against developers to assess its efficacy with practitioners, and conduct a heuristic evaluation against three popular computer vision services. We determine that developers demand certain documentation artefacts more than others, since not all documentation artefacts are equally valued. We map the value (to developers) of these artefacts against their exposure within the software engineering literature, thereby highlighting the gaps by which future research could expand upon. Furthermore, we present a similar mapping against how well the coverage computer vision services have incorporated such artefacts into their own API documentation, thus highlighting that while industry vendors cover most documentation artefacts that may not be in the interest to researchers, some artefacts with low research interest are still largely missing (see \cref{tse2020:tab:high-ips-low-ils-low-cvs}). We therefore provide several generalised recommendations to vendors and the wider research community to explore how best these artefacts can be better addressed and incorporated into further research, thus improving our understanding of the requirements of good API documentation.

Future extensions of our work may involve a restricted systematic literature review in API documentation artefacts, and many suggestions are further detailed in \cref{tse2020:sec:limitations}. Further, a review into the techniques of these primary studies may extend the mapping we conducted in this work, by evaluating the the effectiveness of the various approaches used in each study and assessing these against the proposed conclusions of each study.

The findings of our work provides a solid baseline for improving the documentation of non-deterministic software, such as computer vision services. While our aim is to eventually improve the quality of API documentation, the ultimate goal is to improve the software engineer's experience of non-deterministic and abstracted AI-based components, such as intelligent web services. We hope the guidelines from this extensive study help both software developers and API providers alike by using our taxonomy as a go-to checklist for what should be considered in documenting any API.

\section*{Acknowledgements}

Cummaudo is supported by an Australian Government Research Training Program (RTP) Scholarship.
Grundy is supported by Laureate Fellowship FL190100035. 
We acknowledge additional support provided by the ARC Research Hub for Digital Enhanced Living IH170100013.

\footnotesize
\bibliography{references}

\begin{thebibliography}{43}
\expandafter\ifx\csname natexlab\endcsname\relax\def\natexlab#1{#1}\fi
\providecommand{\bibinfo}[2]{#2}
\ifx\xfnm\relax \def\xfnm[#1]{\unskip,\space#1}\fi
\bibitem[{{Google LLC}(2018)}]{GoogleCloud:Home}
\bibinfo{author}{{Google LLC}}, \bibinfo{title}{{Vision API - Image Content
  Analysis  |  Cloud Vision API  |  Google Cloud}},
  \bibinfo{howpublished}{\url{http://bit.ly/2TD9mBs}}, \bibinfo{year}{2018}.
  \bibinfo{note}{Accessed: 13 September 2018}.
\bibitem[{{Amazon Web Services, Inc.}(2018)}]{AWS:Home}
\bibinfo{author}{{Amazon Web Services, Inc.}}, \bibinfo{title}{{Amazon
  Rekognition}}, \bibinfo{howpublished}{\url{https://amzn.to/2TyT2BL}},
  \bibinfo{year}{2018}. \bibinfo{note}{Accessed: 13 September 2018}.
\bibitem[{{Microsoft Corporation}(2018)}]{Azure:Home}
\bibinfo{author}{{Microsoft Corporation}}, \bibinfo{title}{{Image Processing
  with the Computer Vision API | Microsoft Azure}},
  \bibinfo{howpublished}{\url{http://bit.ly/2YqhkS6}}, \bibinfo{year}{2018}.
  \bibinfo{note}{Accessed: 13 September 2018}.
\bibitem[{{International Business Machines Corporation}(2018)}]{IBM:Home}
\bibinfo{author}{{International Business Machines Corporation}},
  \bibinfo{title}{{Watson Visual Recognition - Overview | IBM}},
  \bibinfo{howpublished}{\url{https://ibm.co/2TBNIO4}}, \bibinfo{year}{2018}.
  \bibinfo{note}{Accessed: 13 September 2018}.
\bibitem[{{Symisc Systems, S.U.A.R.L}(2018)}]{Pixlab:Home}
\bibinfo{author}{{Symisc Systems, S.U.A.R.L}}, \bibinfo{title}{{Computer Vision
  \& Media Processing APIs | PixLab}},
  \bibinfo{howpublished}{\url{http://bit.ly/2UlkW9K}}, \bibinfo{year}{2018}.
  \bibinfo{note}{Accessed: 13 September 2018}.
\bibitem[{{Clarifai, Inc.}(2018)}]{Clarifai:Home}
\bibinfo{author}{{Clarifai, Inc.}}, \bibinfo{title}{{Enterprise AI Powered
  Computer Vision Solutions | Clarifai}},
  \bibinfo{howpublished}{\url{http://bit.ly/2TB3kSa}}, \bibinfo{year}{2018}.
  \bibinfo{note}{Accessed: 13 September 2018}.
\bibitem[{{CloudSight, Inc.}(2018)}]{Cloudsight:Home}
\bibinfo{author}{{CloudSight, Inc.}}, \bibinfo{title}{{Image Recognition API \&
  Visual Search Results | CloudSight AI}},
  \bibinfo{howpublished}{\url{http://bit.ly/2UmNPCw}}, \bibinfo{year}{2018}.
  \bibinfo{note}{Accessed: 13 September 2018}.
\bibitem[{{Deep AI, Inc.}(2018)}]{DeepAI:Home}
\bibinfo{author}{{Deep AI, Inc.}}, \bibinfo{title}{{DeepAI: The front page of
  A.I. | DeepAI}}, \bibinfo{howpublished}{\url{http://bit.ly/2TBNYgf}},
  \bibinfo{year}{2018}. \bibinfo{note}{Accessed: 26 September 2018}.
\bibitem[{{Imagga Technologies}(2018)}]{Imagaa:Home}
\bibinfo{author}{{Imagga Technologies}}, \bibinfo{title}{{Imagga - powerful
  image recognition APIs for automated categorization \& tagging in the cloud
  and on-premises}}, \bibinfo{howpublished}{\url{http://bit.ly/2TxsyRe}},
  \bibinfo{year}{2018}. \bibinfo{note}{Accessed: 13 September 2018}.
\bibitem[{{Talkwalker Inc.}(2018)}]{Talkwaler:Home}
\bibinfo{author}{{Talkwalker Inc.}}, \bibinfo{title}{{Image Recognition -
  Talkwalker}}, \bibinfo{howpublished}{\url{http://bit.ly/2TyT7W5}},
  \bibinfo{year}{2018}. \bibinfo{note}{Accessed: 13 September 2018}.
\bibitem[{{Beijing Kuangshi Technology Co., Ltd.}(2018)}]{Megvii:Home}
\bibinfo{author}{{Beijing Kuangshi Technology Co., Ltd.}},
  \bibinfo{title}{{Megvii}},
  \bibinfo{howpublished}{\url{http://bit.ly/2WJYFzk}}, \bibinfo{year}{2018}.
  \bibinfo{note}{Accessed: 3 April 2019}.
\bibitem[{{Guangzhou Tup Network Technology}(2018)}]{TupuTech:Home}
\bibinfo{author}{{Guangzhou Tup Network Technology}},
  \bibinfo{title}{{TupuTech}},
  \bibinfo{howpublished}{\url{http://bit.ly/2uF4IsN}}, \bibinfo{year}{2018}.
  \bibinfo{note}{Accessed: 3 April 2019}.
\bibitem[{{Shanghai Yitu Technology Co., Ltd.}(2018)}]{YiTuTech:Home}
\bibinfo{author}{{Shanghai Yitu Technology Co., Ltd.}}, \bibinfo{title}{{Yitu
  Technology}}, \bibinfo{howpublished}{\url{http://bit.ly/2uGvxgf}},
  \bibinfo{year}{2018}. \bibinfo{note}{Accessed: 3 April 2019}.
\bibitem[{{SenseTime}(2018)}]{SenseTime:Home}
\bibinfo{author}{{SenseTime}}, \bibinfo{title}{{SenseTime}},
  \bibinfo{howpublished}{\url{http://bit.ly/2WH6RjF}}, \bibinfo{year}{2018}.
  \bibinfo{note}{Accessed: 3 April 2019}.
\bibitem[{{Beijing Geling Shentong Information Technology Co.,
  Ltd.}(2018)}]{DeepGlint:Home}
\bibinfo{author}{{Beijing Geling Shentong Information Technology Co., Ltd.}},
  \bibinfo{title}{{DeepGlint}},
  \bibinfo{howpublished}{\url{http://bit.ly/2uHHdPS}}, \bibinfo{year}{2018}.
  \bibinfo{note}{Accessed: 3 April 2019}.
\bibitem[{{Google LLC}(2019)}]{Quicksta92:online}
\bibinfo{author}{{Google LLC}}, \bibinfo{title}{Quickstart: Using client
  libraries  |  cloud vision api documentation  |  google cloud},
  \bibinfo{howpublished}{\url{http://bit.ly/2RRMQHG}}, \bibinfo{year}{2019}.
  \bibinfo{note}{Accessed: 18 March 2019}.
\bibitem[{{Microsoft Corporation}(2019)}]{CalltheC0:online}
\bibinfo{author}{{Microsoft Corporation}}, \bibinfo{title}{Call the computer
  vision api - azure cognitive services | microsoft docs},
  \bibinfo{howpublished}{\url{http://bit.ly/2vHSdjT}}, \bibinfo{year}{2019}.
  \bibinfo{note}{Accessed: 18 March 2019}.
\bibitem[{{Amazon Web Services, Inc.}(2019)}]{Troubles2:online}
\bibinfo{author}{{Amazon Web Services, Inc.}}, \bibinfo{title}{Troubleshooting
  amazon rekognition video - amazon rekognition},
  \bibinfo{howpublished}{\url{https://amzn.to/3b763fS}}, \bibinfo{year}{2019}.
  \bibinfo{note}{Accessed: 18 March 2019}.
\bibitem[{{Microsoft Corporation}(2019{\natexlab{a}})}]{LegalTermsMS:online}
\bibinfo{author}{{Microsoft Corporation}}, \bibinfo{title}{Microsoft azure
  legal information | microsoft azure},
  \bibinfo{howpublished}{\url{https://bit.ly/2Cy8Z8r}},
  \bibinfo{year}{2019}{\natexlab{a}}. \bibinfo{note}{Accessed: 18 March 2019}.
\bibitem[{{Microsoft Corporation}(2019{\natexlab{b}})}]{Quicksta70:online}
\bibinfo{author}{{Microsoft Corporation}}, \bibinfo{title}{Quickstart: Computer
  vision client library for .net - azure cognitive services | microsoft docs},
  \bibinfo{howpublished}{\url{http://bit.ly/2vF3wJC}},
  \bibinfo{year}{2019}{\natexlab{b}}. \bibinfo{note}{Accessed: 18 March 2019}.
\bibitem[{{Amazon Web Services, Inc.}(2019{\natexlab{a}})}]{Exercise86:online}
\bibinfo{author}{{Amazon Web Services, Inc.}}, \bibinfo{title}{Exercise 1:
  Detect objects and scenes in an image (console) - amazon rekognition},
  \bibinfo{howpublished}{\url{https://amzn.to/36TkLnm}},
  \bibinfo{year}{2019}{\natexlab{a}}. \bibinfo{note}{Accessed: 18 March 2019}.
\bibitem[{{Amazon Web Services, Inc.}(2019{\natexlab{b}})}]{AmazonRe55:online}
\bibinfo{author}{{Amazon Web Services, Inc.}}, \bibinfo{title}{Amazon
  rekognition | aws machine learning blog},
  \bibinfo{howpublished}{\url{https://go.aws/37Q7lKc}},
  \bibinfo{year}{2019}{\natexlab{b}}. \bibinfo{note}{Accessed: 18 March 2019}.
\bibitem[{{Microsoft Corporation}(2019{\natexlab{a}})}]{Tutorial42:online}
\bibinfo{author}{{Microsoft Corporation}}, \bibinfo{title}{Tutorial: Generate
  metadata for azure images - azure cognitive services | microsoft docs},
  \bibinfo{howpublished}{\url{http://bit.ly/2RRnARK}},
  \bibinfo{year}{2019}{\natexlab{a}}. \bibinfo{note}{Accessed: 18 March 2019}.
\bibitem[{{Microsoft Corporation}(2019{\natexlab{b}})}]{Tutorial6:online}
\bibinfo{author}{{Microsoft Corporation}}, \bibinfo{title}{Tutorial: Use custom
  logo detector to recognize azure services - custom vision - azure cognitive
  services | microsoft docs},
  \bibinfo{howpublished}{\url{http://bit.ly/2RUGwPH}},
  \bibinfo{year}{2019}{\natexlab{b}}. \bibinfo{note}{Accessed: 18 March 2019}.
\bibitem[{{Microsoft Corporation}(2019{\natexlab{c}})}]{GitHubAz56:online}
\bibinfo{author}{{Microsoft Corporation}}, \bibinfo{title}{Github -
  azure-samples/cognitive-services-java-computer-vision-tutorial: This tutorial
  shows the features of the microsoft cognitive services computer vision rest
  api.}, \bibinfo{howpublished}{\url{http://bit.ly/37N1yoN}},
  \bibinfo{year}{2019}{\natexlab{c}}. \bibinfo{note}{Accessed: 18 March 2019}.
\bibitem[{{Microsoft Corporation}(2019{\natexlab{d}})}]{SampleEx87:online}
\bibinfo{author}{{Microsoft Corporation}}, \bibinfo{title}{Sample: Explore an
  image processing app in c\# - azure cognitive services | microsoft docs},
  \bibinfo{howpublished}{\url{http://bit.ly/2u4mPMh}},
  \bibinfo{year}{2019}{\natexlab{d}}. \bibinfo{note}{Accessed: 18 March 2019}.
\bibitem[{{Amazon Web Services, Inc.}(2019)}]{JavaSDKV25:online}
\bibinfo{author}{{Amazon Web Services, Inc.}}, \bibinfo{title}{Java (sdk v1)
  code samples for amazon rekognition - aws code sample},
  \bibinfo{howpublished}{\url{https://amzn.to/2ugTle3}}, \bibinfo{year}{2019}.
  \bibinfo{note}{Accessed: 18 March 2019}.
\bibitem[{{Google LLC}(2019{\natexlab{a}})}]{SampleAp41:online}
\bibinfo{author}{{Google LLC}}, \bibinfo{title}{Sample applications  |  cloud
  vision api documentation  |  google cloud},
  \bibinfo{howpublished}{\url{http://bit.ly/2SdoB5r}},
  \bibinfo{year}{2019}{\natexlab{a}}. \bibinfo{note}{Accessed: 18 March 2019}.
\bibitem[{{Google LLC}(2019{\natexlab{b}})}]{Bestprac23:online}
\bibinfo{author}{{Google LLC}}, \bibinfo{title}{Best practices for enterprise
  organizations  |  documentation  |  google cloud},
  \bibinfo{howpublished}{\url{http://bit.ly/2v0RSs5}},
  \bibinfo{year}{2019}{\natexlab{b}}. \bibinfo{note}{Accessed: 18 March 2019}.
\bibitem[{{Google LLC}(2019{\natexlab{c}})}]{TipsTric26:online}
\bibinfo{author}{{Google LLC}}, \bibinfo{title}{Tips \& tricks  |  cloud
  functions documentation  |  google cloud},
  \bibinfo{howpublished}{\url{http://bit.ly/2GZNc8Z}},
  \bibinfo{year}{2019}{\natexlab{c}}. \bibinfo{note}{Accessed: 18 March 2019}.
\bibitem[{{Microsoft Corporation}(2019)}]{Improvin23:online}
\bibinfo{author}{{Microsoft Corporation}}, \bibinfo{title}{Improving your
  classifier - custom vision service - azure cognitive services | microsoft
  docs}, \bibinfo{howpublished}{\url{http://bit.ly/37SBkRQ}},
  \bibinfo{year}{2019}. \bibinfo{note}{Accessed: 18 March 2019}.
\bibitem[{{Amazon Web Services, Inc.}(2019{\natexlab{a}})}]{BestPrac58:online}
\bibinfo{author}{{Amazon Web Services, Inc.}}, \bibinfo{title}{Best practices
  for sensors, input images, and videos - amazon rekognition},
  \bibinfo{howpublished}{\url{https://amzn.to/2uZlWo0}},
  \bibinfo{year}{2019}{\natexlab{a}}. \bibinfo{note}{Accessed: 18 March 2019}.
\bibitem[{{Amazon Web Services, Inc.}(2019{\natexlab{b}})}]{AmazonRe70:online}
\bibinfo{author}{{Amazon Web Services, Inc.}}, \bibinfo{title}{Amazon
  rekognition image}, \bibinfo{howpublished}{\url{https://go.aws/2ubB6qc}},
  \bibinfo{year}{2019}{\natexlab{b}}. \bibinfo{note}{Accessed: 18 March 2019}.
\bibitem[{{Amazon Web Services, Inc.}(2019{\natexlab{c}})}]{ActionsA39:online}
\bibinfo{author}{{Amazon Web Services, Inc.}}, \bibinfo{title}{Actions - amazon
  rekognition}, \bibinfo{howpublished}{\url{https://amzn.to/392p3dH}},
  \bibinfo{year}{2019}{\natexlab{c}}. \bibinfo{note}{Accessed: 18 March 2019}.
\bibitem[{{Microsoft Corporation}(2019)}]{WhatisCo90:online}
\bibinfo{author}{{Microsoft Corporation}}, \bibinfo{title}{What is computer
  vision? - computer vision - azure cognitive services | microsoft docs},
  \bibinfo{howpublished}{\url{http://bit.ly/37SomDx}}, \bibinfo{year}{2019}.
  \bibinfo{note}{Accessed: 18 March 2019}.
\bibitem[{{Google LLC}(2019)}]{VisionAI32:online}
\bibinfo{author}{{Google LLC}}, \bibinfo{title}{Vision ai | derive image
  insights via ml  |  cloud vision api  |  google cloud},
  \bibinfo{howpublished}{\url{http://bit.ly/31nWoNx}}, \bibinfo{year}{2019}.
  \bibinfo{note}{Accessed: 18 March 2019}.
\bibitem[{{Amazon Web Services, Inc.}(2019)}]{AWSRelea46:online}
\bibinfo{author}{{Amazon Web Services, Inc.}}, \bibinfo{title}{Aws release
  notes}, \bibinfo{howpublished}{\url{https://go.aws/2v0RYjr}},
  \bibinfo{year}{2019}. \bibinfo{note}{Accessed: 18 March 2019}.
\bibitem[{{Google LLC}(2019)}]{ReleaseN91:online}
\bibinfo{author}{{Google LLC}}, \bibinfo{title}{Release notes  |  cloud
  vision api documentation  |  google cloud},
  \bibinfo{howpublished}{\url{http://bit.ly/2UipY5J}}, \bibinfo{year}{2019}.
  \bibinfo{note}{Accessed: 18 March 2019}.
\bibitem[{{Microsoft Corporation}(2019)}]{ReleaseN4:online}
\bibinfo{author}{{Microsoft Corporation}}, \bibinfo{title}{Release notes -
  custom vision service - azure cognitive services | microsoft docs},
  \bibinfo{howpublished}{\url{http://bit.ly/2UlPiaW}}, \bibinfo{year}{2019}.
  \bibinfo{note}{Accessed: 18 March 2019}.
\bibitem[{{Amazon Web Services, Inc.}(2019{\natexlab{a}})}]{Step1Set76:online}
\bibinfo{author}{{Amazon Web Services, Inc.}}, \bibinfo{title}{Step 1: Set up
  an aws account and create an iam user - amazon rekognition},
  \bibinfo{howpublished}{\url{https://amzn.to/2tqW4kI}},
  \bibinfo{year}{2019}{\natexlab{a}}. \bibinfo{note}{Accessed: 18 March 2019}.
\bibitem[{{Amazon Web Services, Inc.}(2019{\natexlab{b}})}]{Limitsin66:online}
\bibinfo{author}{{Amazon Web Services, Inc.}}, \bibinfo{title}{Limits in amazon
  rekognition - amazon rekognition},
  \bibinfo{howpublished}{\url{https://amzn.to/2On6n0h}},
  \bibinfo{year}{2019}{\natexlab{b}}. \bibinfo{note}{Accessed: 18 March 2019}.
\bibitem[{{Microsoft Corporation}(2019)}]{Contentt49:online}
\bibinfo{author}{{Microsoft Corporation}}, \bibinfo{title}{Content tags -
  computer vision - azure cognitive services | microsoft docs},
  \bibinfo{howpublished}{\url{http://bit.ly/2vESzHX}}, \bibinfo{year}{2019}.
  \bibinfo{note}{Accessed: 18 March 2019}.
\bibitem[{{Google LLC}(2019)}]{MachineL36:online}
\bibinfo{author}{{Google LLC}}, \bibinfo{title}{Machine learning glossary  | 
  google developers}, \bibinfo{howpublished}{\url{http://bit.ly/3b38VdL}},
  \bibinfo{year}{2019}. \bibinfo{note}{Accessed: 18 March 2019}.

\end{thebibliography}


\begin{thebibliography}{59}
\providecommand{\natexlab}[1]{#1}
\providecommand{\url}[1]{#1}
\csname url@samestyle\endcsname
\providecommand{\newblock}{\relax}
\providecommand{\bibinfo}[2]{#2}
\providecommand{\BIBentrySTDinterwordspacing}{\spaceskip=0pt\relax}
\providecommand{\BIBentryALTinterwordstretchfactor}{4}
\providecommand{\BIBentryALTinterwordspacing}{\spaceskip=\fontdimen2\font plus
\BIBentryALTinterwordstretchfactor\fontdimen3\font minus
  \fontdimen4\font\relax}
\providecommand{\BIBforeignlanguage}[2]{{%
\expandafter\ifx\csname l@#1\endcsname\relax
\typeout{** WARNING: IEEEtranSN.bst: No hyphenation pattern has been}%
\typeout{** loaded for the language `#1'. Using the pattern for}%
\typeout{** the default language instead.}%
\else
\language=\csname l@#1\endcsname
\fi
#2}}
\providecommand{\BIBdecl}{\relax}
\BIBdecl

\bibitem[Aghajani et~al.(2019)Aghajani, Nagy, Vega-Marquez, Linares-Vasquez,
  Moreno, Bavota, and Lanza]{Aghajani:2019bo}
E.~Aghajani, C.~Nagy, O.~L. Vega-Marquez, M.~Linares-Vasquez, L.~Moreno,
  G.~Bavota, and M.~Lanza, ``{Software Documentation Issues Unveiled},'' in
  \emph{Proceedings of the 41st International Conference on Software
  Engineering}.\hskip 1em plus 0.5em minus 0.4em\relax Montreal, QC, Canada:
  IEEE, May 2019, pp. 1199--1210.

\bibitem[Aversano et~al.(2017)Aversano, Guardabascio, and
  Tortorella]{Aversano:2017ic}
L.~Aversano, D.~Guardabascio, and M.~Tortorella, ``{Analysis of the
  Documentation of ERP Software Projects},'' \emph{Procedia Computer Science},
  vol. 121, pp. 423--430, January 2017.

\bibitem[Baruch(1999)]{Baruch:1999vf}
Y.~Baruch, ``{Response rate in academic studies - A comparative analysis},''
  \emph{Human Relations}, vol.~52, no.~4, pp. 421--438, 1999.

\bibitem[Bottomley(2005)]{Bottomley:2005fs}
C.~Bottomley, ``{What part writer? What part programmer? A survey of practices
  and knowledge used in programmer writing},'' in \emph{Proceedings of the 2005
  IEEE International Professional Communication Conference}.\hskip 1em plus
  0.5em minus 0.4em\relax Limerick, Ireland: IEEE, July 2005, pp. 802--812.

\bibitem[Brereton et~al.(2007)Brereton, Kitchenham, Budgen, Turner, and
  Khalil]{Brereton:2007by}
P.~Brereton, B.~A. Kitchenham, D.~Budgen, M.~Turner, and M.~Khalil, ``{Lessons
  from applying the systematic literature review process within the software
  engineering domain},'' \emph{Journal of Systems and Software}, vol.~80,
  no.~4, pp. 571--583, April 2007.

\bibitem[Brooke(1996)]{Brooke:1996ua}
J.~Brooke, ``{SUS-A quick and dirty usability scale},'' \emph{Usability
  Evaluation in Industry}, pp. 189--194, 1996.

\bibitem[Brooke(2013)]{Brooke:2013vt}
------, ``{SUS: a retrospective},'' \emph{Journal of Usability Studies},
  vol.~8, no.~2, pp. 29--40, 2013.

\bibitem[Cicchetti(1994)]{cicchetti1994guidelines}
D.~V. Cicchetti, ``{Guidelines, Criteria, and Rules of Thumb for Evaluating
  Normed and Standardized Assessment Instruments in Psychology},''
  \emph{Psychological Assessment}, vol.~6, no.~4, pp. 284--290, 1994.

\bibitem[Cummaudo et~al.(2019{\natexlab{b}})Cummaudo, Vasa, and
  Grundy]{Cummaudo:2019esem}
A.~Cummaudo, R.~Vasa, and J.~Grundy, ``{What should I document? A preliminary
  systematic mapping study into API documentation knowledge},'' in
  \emph{Proceedings of the 13th International Symposium on Empirical Software
  Engineering and Measurement}.\hskip 1em plus 0.5em minus 0.4em\relax Porto de
  Galinhas, Recife, Brazil: IEEE, October 2019, pp. 1--6.

\bibitem[Cummaudo et~al.(2019{\natexlab{a}})Cummaudo, Vasa, Grundy, Abdelrazek,
  and Cain]{Cummaudo:2019icsme}
A.~Cummaudo, R.~Vasa, J.~Grundy, M.~Abdelrazek, and A.~Cain, ``{Losing
  Confidence in Quality: Unspoken Evolution of Computer Vision Services},'' in
  \emph{Proceedings of the 35th IEEE International Conference on Software
  Maintenance and Evolution}.\hskip 1em plus 0.5em minus 0.4em\relax Cleveland,
  OH, USA: IEEE, December 2019, pp. 333--342.

\bibitem[Cummaudo et~al.(2020)Cummaudo, Vasa, Barnett, Grundy, and
  Abdelrazek]{Cummaudo:2020icse}
A.~Cummaudo, R.~Vasa, S.~Barnett, J.~Grundy, and M.~Abdelrazek, ``{Interpreting
  Cloud Computer Vision Pain-Points: A Mining Study of Stack Overflow},'' in
  \emph{Proceedings of the 42nd International Conference on Software
  Engineering}.\hskip 1em plus 0.5em minus 0.4em\relax Seoul, Republic of
  Korea: IEEE, May 2020, {In Press}.

\bibitem[Dey(1993)]{Dey:2003ty}
I.~Dey, \emph{{Qualitative Data Analysis: A User-Friendly Guide for Social
  Scientists}}.\hskip 1em plus 0.5em minus 0.4em\relax New York, NY: Routledge,
  1993.

\bibitem[Gamer et~al.(2010)Gamer, Lemon, Fellows, and Singh]{Gamer:tj}
M.~Gamer, J.~Lemon, I.~Fellows, and P.~Singh, ``{Irr: various coefficients of
  interrater reliability},'' \emph{R package version 0.83}, 2010.

\bibitem[Garousi and Felderer(2017)]{Garousi:2017:EGE:3084226.3084238}
V.~Garousi and M.~Felderer, ``{Experience-based guidelines for effective and
  efficient data extraction in systematic reviews in software engineering},''
  in \emph{Proceedings of the 21st International Conference on Evaluation and
  Assessment in Software Engineering}, vol. Part F1286.\hskip 1em plus 0.5em
  minus 0.4em\relax Karlskrona, Sweden: ACM, June 2017, pp. 170--179.

\bibitem[Garousi et~al.(2019)Garousi, Felderer, and
  M{\"{a}}ntyl{\"{a}}]{GAROUSI2019101}
V.~Garousi, M.~Felderer, and M.~V. M{\"{a}}ntyl{\"{a}}, ``{Guidelines for
  including grey literature and conducting multivocal literature reviews in
  software engineering},'' \emph{Information and Software Technology}, vol.
  106, pp. 101--121, 2019.

\bibitem[Geiger et~al.(2018)Geiger, Varoquaux, Mazel-Cabasse, and
  Holdgraf]{Geiger:2018fv}
R.~S. Geiger, N.~Varoquaux, C.~Mazel-Cabasse, and C.~Holdgraf, ``{The Types,
  Roles, and Practices of Documentation in Data Analytics Open Source Software
  Libraries: A Collaborative Ethnography of Documentation Work},''
  \emph{Computer Supported Cooperative Work: CSCW: An International Journal},
  vol.~27, no. 3-6, pp. 767--802, May 2018.

\bibitem[Glass et~al.(2009)Glass, Vessey, and Ramesh]{Glass:2002wa}
R.~L. Glass, I.~Vessey, and V.~Ramesh, ``{RESRES: The story behind the paper
  "Research in software engineering: An analysis of the literature"},''
  \emph{Information and Software Technology}, vol.~51, no.~1, pp. 68--70, 2009.

\bibitem[Hallgren(2012)]{Hallgren:2012kt}
K.~A. Hallgren, ``{Computing Inter-Rater Reliability for Observational Data: An
  Overview and Tutorial},'' \emph{Tutorials in Quantitative Methods for
  Psychology}, vol.~8, no.~1, pp. 23--34, February 2012.

\bibitem[Haselbock et~al.(2019)Haselbock, Weinreich, Buchgeher, and
  Kriechbaum]{Haselbock:2018jd}
S.~Haselbock, R.~Weinreich, G.~Buchgeher, and T.~Kriechbaum, ``{Microservice
  Design Space Analysis and Decision Documentation: A Case Study on API
  Management},'' in \emph{Proceedings of the 11th International Conference on
  Service-Oriented Computing and Applications, SOCA 2018}, Paris, France,
  November 2019, pp. 1--8.

\bibitem[Head et~al.(2018)Head, Sadowski, Murphy-Hill, and
  Knight]{Head:2018baa}
A.~Head, C.~Sadowski, E.~Murphy-Hill, and A.~Knight, ``{When not to comment:
  Questions and tradeoffs with API documentation for C++ projects},'' in
  \emph{Proceedings of the 40th International Conference on Software
  Engineering}, ser. questions and tradeoffs with API documentation for C++
  projects.\hskip 1em plus 0.5em minus 0.4em\relax Gothenburg, Sweden: ACM, May
  2018, pp. 643--653.

\bibitem[Hosseini et~al.(2017)Hosseini, Xiao, and Poovendran]{Hosseini:2018jr}
H.~Hosseini, B.~Xiao, and R.~Poovendran, ``{Google's cloud vision API is not
  robust to noise},'' in \emph{Proceedings of the 16th IEEE International
  Conference on Machine Learning and Applications}, vol. 2017-Decem.\hskip 1em
  plus 0.5em minus 0.4em\relax Cancun, Mexico: IEEE, December 2017, pp.
  101--105.

\bibitem[IEEE(1990)]{IEEE:1990wp}
IEEE, ``{IEEE Standard Glossary of Software Engineering Terminology},'' 1990.

\bibitem[Inzunza et~al.(2018)Inzunza, Ju{\'{a}}rez-Ram{\'{i}}rez, and
  Jim{\'{e}}nez]{Inzunza:2018dn}
S.~Inzunza, R.~Ju{\'{a}}rez-Ram{\'{i}}rez, and S.~Jim{\'{e}}nez, ``{API
  Documentation},'' in \emph{Proceedings of the 6th World Conference on
  Information Systems and Technologies}.\hskip 1em plus 0.5em minus 0.4em\relax
  Naples, Italy: Springer, March 2018, pp. 229--239.

\bibitem[Jeong et~al.(2009)Jeong, Xie, Beaton, Myers, Stylos, Ehret, Karstens,
  Efeoglu, and Busse]{Jeong:2009tu}
S.~Y. Jeong, Y.~Xie, J.~Beaton, B.~A. Myers, J.~Stylos, R.~Ehret, J.~Karstens,
  A.~Efeoglu, and D.~K. Busse, ``{Improving documentation for eSOA APIs through
  user studies},'' in \emph{Proceedings of the First International Symposium on
  End User Development}, vol. 5435 LNCS.\hskip 1em plus 0.5em minus 0.4em\relax
  Siegen, Germany: Springer, March 2009, pp. 86--105.

\bibitem[Jiarpakdee et~al.(2020)Jiarpakdee, Tantithamthavorn, Dam, and
  Grundy]{Jiarpakdee2020}
J.~Jiarpakdee, C.~Tantithamthavorn, H.~K. Dam, and J.~Grundy, ``{An Empirical
  Study of Model-Agnostic Techniques for Defect Prediction Models},''
  \emph{IEEE Transactions on Software Engineering}, vol. 5589, no.~c, pp. 1--1,
  2020.

\bibitem[Kitchenham and Charters(2007)]{Kitchenham:2007dd}
B.~Kitchenham and S.~Charters, ``{Guidelines for performing Systematic
  Literature Reviews in Software Engineering},'' Software Engineering Group,
  Keele University and Department of Computer Science, University of Durham,
  Keele, UK, Tech. Rep., 2007.

\bibitem[Kitchenham and Pfleeger(2007)]{Kitchenham:2007ux}
B.~A. Kitchenham and S.~L. Pfleeger, ``{Personal opinion surveys},'' in
  \emph{Guide to Advanced Empirical Software Engineering}, F.~Shull, J.~Singer,
  and D.~I.~K. Sj{\o}berg, Eds.\hskip 1em plus 0.5em minus 0.4em\relax
  Springer, November 2007, ch.~3, pp. 63--92.

\bibitem[Ko and Riche(2011)]{Ko:2011fb}
A.~J. Ko and Y.~Riche, ``{The role of conceptual knowledge in API usability},''
  in \emph{Proceedings of the 2011 IEEE Symposium on Visual Languages and Human
  Centric Computing}.\hskip 1em plus 0.5em minus 0.4em\relax Pittsburg, PA,
  USA: IEEE, September 2011, pp. 173--176.

\bibitem[Kotula(1998)]{Kotula:1998wp}
J.~Kotula, ``{Using patterns to create component documentation},'' \emph{IEEE
  Software}, vol.~15, no.~2, pp. 84--92, 1998.

\bibitem[Krosnick(1999)]{Krosnick:1999wt}
J.~A. Krosnick, ``{Survey Research},'' \emph{Annual Review of Psychology},
  vol.~50, no.~1, pp. 537--567, February 1999.

\bibitem[Landis and Koch(1977)]{Landis:1977kv}
J.~R. Landis and G.~G. Koch, ``{The Measurement of Observer Agreement for
  Categorical Data},'' \emph{Biometrics}, vol.~33, no.~1, p. 159, March 1977.

\bibitem[Lethbridge et~al.(2005)Lethbridge, Sim, and Singer]{Lethbridge:2005jv}
T.~C. Lethbridge, S.~E. Sim, and J.~Singer, ``{Studying software engineers:
  Data collection techniques for software field studies},'' \emph{Empirical
  Software Engineering}, vol.~10, no.~3, pp. 311--341, July 2005.

\bibitem[Maalej and Robillard(2013)]{Maalej2013}
W.~Maalej and M.~P. Robillard, ``{Patterns of knowledge in API reference
  documentation},'' \emph{IEEE Transactions on Software Engineering}, 2013.

\bibitem[McLellan et~al.(1998)McLellan, Roesler, Tempest, and
  Spinuzzi]{McLellan:1998vu}
S.~G. McLellan, A.~W. Roesler, J.~T. Tempest, and C.~I. Spinuzzi, ``{Building
  more usable APIs},'' \emph{IEEE Software}, vol.~15, no.~3, pp. 78--86, 1998.

\bibitem[McLeod and MacDonell(2011)]{mcleod2011factors}
L.~McLeod and S.~G. MacDonell, ``{Factors that affect software systems
  development project outcomes: A survey of research},'' \emph{ACM Computing
  Surveys}, vol.~43, no.~4, p.~24, 2011.

\bibitem[Meng et~al.(2018)Meng, Steinhardt, and Schubert]{Meng:2017cx}
M.~Meng, S.~Steinhardt, and A.~Schubert, ``{Application programming interface
  documentation: What do software developers want?}'' \emph{Journal of
  Technical Writing and Communication}, vol.~48, no.~3, pp. 295--330, August
  2018.

\bibitem[{Myers} et~al.(2016){Myers}, {Ko}, {LaToza}, and {Yoon}]{7503516}
B.~A. {Myers}, A.~J. {Ko}, T.~D. {LaToza}, and Y.~{Yoon}, ``Programmers are
  users too: Human-centered methods for improving programming tools,''
  \emph{Computer}, vol.~49, no.~7, pp. 44--52, 2016.

\bibitem[Myers and Stylos(2016)]{myersstylos2016}
B.~A. Myers and J.~Stylos, ``{Improving API Usability},'' \emph{Communications
  of the ACM}, vol.~59, no.~6, p. 62–69, May 2016.

\bibitem[Nybom et~al.(2018)Nybom, Ashraf, and Porres]{Nybom:2018ef}
K.~Nybom, A.~Ashraf, and I.~Porres, ``{A systematic mapping study on API
  documentation generation approaches},'' in \emph{Proceedings of the 44th
  Euromicro Conference on Software Engineering and Advanced
  Applications}.\hskip 1em plus 0.5em minus 0.4em\relax Prague, Czech Republic:
  IEEE, August 2018, pp. 462--469.

\bibitem[Nykaza et~al.(2002)Nykaza, Messinger, Boehme, Norman, Mace, and
  Gordon]{Nykaza:2002td}
J.~Nykaza, R.~Messinger, F.~Boehme, C.~L. Norman, M.~Mace, and M.~Gordon,
  ``{What programmers really want: Results of a needs assessment for SDK
  documentation},'' in \emph{Proceedings of the 20th Annual International
  Conference on Computer Documentation}.\hskip 1em plus 0.5em minus 0.4em\relax
  Toronto, ON, Canada: ACM, October 2002, pp. 133--141.

\bibitem[Parnas and Vilkomir(2007)]{Parnas:2007fb}
D.~L. Parnas and S.~A. Vilkomir, ``{Precise documentation of critical
  software},'' in \emph{Proceedings of 10th IEEE International Symposium on
  High Assurance Systems Engineering}.\hskip 1em plus 0.5em minus 0.4em\relax
  Plano, TX, USA: IEEE, November 2007, pp. 237--244.

\bibitem[Petersen et~al.(2008)Petersen, Feldt, Mujtaba, and
  Mattsson]{Petersen:2008td}
K.~Petersen, R.~Feldt, S.~Mujtaba, and M.~Mattsson, ``{Systematic mapping
  studies in software engineering},'' in \emph{Proceedings of the 12th
  International Conference on Evaluation and Assessment in Software
  Engineering, EASE 2008}, 2008, pp. 68--77.

\bibitem[{R Core Team}(2020)]{RCoreTeam}
{R Core Team}, \emph{{R - A Language and Environment for Statistical
  Computing}}, \url{https://www.R-project.org/}, R Foundation for Statistical
  Computing, Vienna, Austria, 2020.

\bibitem[{RightScale Inc.}(2016)]{RightScaleInc:2018kJ}
{RightScale Inc.}, ``{State of the Cloud Report: DevOps Trends},'' Tech. Rep.,
  2016.

\bibitem[Robillard(2009)]{Robillard:2009uk}
M.~P. Robillard, ``{What makes APIs hard to learn? Answers from developers},''
  \emph{IEEE Software}, vol.~26, no.~6, pp. 27--34, 2009.

\bibitem[Robillard and Deline(2011)]{Robillard:2011uv}
M.~P. Robillard and R.~Deline, ``{A field study of API learning obstacles},''
  \emph{Empirical Software Engineering}, vol.~16, no.~6, pp. 703--732, 2011.

\bibitem[Robillard et~al.(2017)Robillard, Marcus, Treude, Bavota, Chaparro,
  Ernst, Gerosall, Godfrey, Lanza, Linares-V{\'{a}}squez, Murphy, Moreno,
  Shepherd, and Wong]{Robillard:hk}
M.~P. Robillard, A.~Marcus, C.~Treude, G.~Bavota, O.~Chaparro, N.~Ernst, M.~A.
  Gerosall, M.~Godfrey, M.~Lanza, M.~Linares-V{\'{a}}squez, G.~C. Murphy,
  L.~Moreno, D.~Shepherd, and E.~Wong, ``{On-demand developer documentation},''
  in \emph{Proceedings of the 33rd IEEE International Conference on Software
  Maintenance and Evolution}.\hskip 1em plus 0.5em minus 0.4em\relax Shanghai,
  China: IEEE, September 2017, pp. 479--483.

\bibitem[Sauro and Lewis(2011)]{Sauro:2011aj}
J.~Sauro and J.~R. Lewis, ``{When designing usability questionnaires, does it
  hurt to be positive?}'' in \emph{Proceedings of the 2011 SIGCHI Conference on
  Human Factors in Computing Systems}, Vancouver, BC, Canada, May 2011, pp.
  2215--2223.

\bibitem[Schwandt(1996)]{Miles:1994ty}
T.~A. Schwandt, ``{Qualitative data analysis: An expanded sourcebook},''
  \emph{Evaluation and Program Planning}, vol.~19, no.~1, pp. 106--107, 1996.

\bibitem[Seaman(2007)]{Seaman:2007wa}
C.~B. Seaman, ``{Qualitative methods},'' in \emph{Guide to Advanced Empirical
  Software Engineering}, F.~Shull, J.~Singer, and D.~I.~K. Sj{\o}berg,
  Eds.\hskip 1em plus 0.5em minus 0.4em\relax Springer, November 2007, ch.~2,
  pp. 35--62.

\bibitem[Singer et~al.(2007)Singer, Sim, and Lethbridge]{Singer:2007tu}
J.~Singer, S.~E. Sim, and T.~C. Lethbridge, ``{Software engineering data
  collection for field studies},'' in \emph{Guide to Advanced Empirical
  Software Engineering}, F.~Shull, J.~Singer, and D.~I.~K. Sj{\o}berg,
  Eds.\hskip 1em plus 0.5em minus 0.4em\relax Springer, November 2007, ch.~1,
  pp. 9--34.

\bibitem[Spector(1992)]{Spector:1992uj}
P.~Spector, \emph{{Summated Rating Scale Construction}}.\hskip 1em plus 0.5em
  minus 0.4em\relax Newbury Park, CA, USA: SAGE, 1992.

\bibitem[Taulavuori et~al.(2004)Taulavuori, Niemel{\"{a}}, and
  Kallio]{Taulavuori:2004el}
A.~Taulavuori, E.~Niemel{\"{a}}, and P.~Kallio, ``{Component documentation - A
  key issue in software product lines},'' \emph{Information and Software
  Technology}, vol.~46, no.~8, pp. 535--546, June 2004.

\bibitem[Usman et~al.(2017)Usman, Britto, B{\"{o}}rstler, and
  Mendes]{Usman:2017hn}
M.~Usman, R.~Britto, J.~B{\"{o}}rstler, and E.~Mendes, ``{Taxonomies in
  software engineering: A Systematic mapping study and a revised taxonomy
  development method},'' \emph{Information and Software Technology}, vol.~85,
  pp. 43--59, May 2017.

\bibitem[Wang et~al.(2018)Wang, Yao, Viswanath, Zheng, and Zhao]{Wang:2018vl}
B.~Wang, Y.~Yao, B.~Viswanath, H.~Zheng, and B.~Y. Zhao, ``{With great training
  comes great vulnerability: Practical attacks against transfer learning},'' in
  \emph{Proceedings of the 27th USENIX Security Symposium}.\hskip 1em plus
  0.5em minus 0.4em\relax Baltimore, MD, USA: USENIX Association, July 2018,
  pp. 1281--1297.

\bibitem[Watson(2012)]{Watson:2012uy}
R.~Watson, ``{Development and application of a heuristic to assess trends in
  API documentation},'' in \emph{Proceedings of the 30th ACM International
  Conference on Design of Communication}.\hskip 1em plus 0.5em minus
  0.4em\relax Seattle, WA, USA: ACM, October 2012, pp. 295--302.

\bibitem[Watson et~al.(2013)Watson, {Mark Stamnes}, Jeannot-Schroeder, and
  Spyridakis]{Watson:2013fx}
R.~Watson, M.~{Mark Stamnes}, J.~Jeannot-Schroeder, and J.~H. Spyridakis,
  ``{API documentation and software community values: A survey of open-source
  API documentation},'' in \emph{Proceedings of the 31st ACM International
  Conference on Design of Communication}.\hskip 1em plus 0.5em minus
  0.4em\relax Greenville, SC, USA: ACM, September 2013, pp. 165--174.

\bibitem[Wieringa et~al.(2006)Wieringa, Maiden, Mead, and
  Rolland]{Wieringa:2006vd}
R.~Wieringa, N.~Maiden, N.~Mead, and C.~Rolland, ``{Requirements engineering
  paper classification and evaluation criteria: a proposal and a discussion},''
  \emph{Requirements Engineering}, vol.~11, no.~1, pp. 102--107, March 2006.

\bibitem[Zhi et~al.(2015)Zhi, Garousi-Yusifoğlu, Sun, Garousi, Shahnewaz, and
  Ruhe]{ZHI2015175}
J.~Zhi, V.~Garousi-Yusifoğlu, B.~Sun, G.~Garousi, S.~Shahnewaz, and G.~Ruhe,
  ``Cost, benefits and quality of software development documentation: A
  systematic mapping,'' \emph{Journal of Systems and Software}, vol.~99, pp.
  175 -- 198, 2015.

\end{thebibliography}

\begin{IEEEbiography}[{\includegraphics[width=1in,height=1.25in,clip,keepaspectratio]{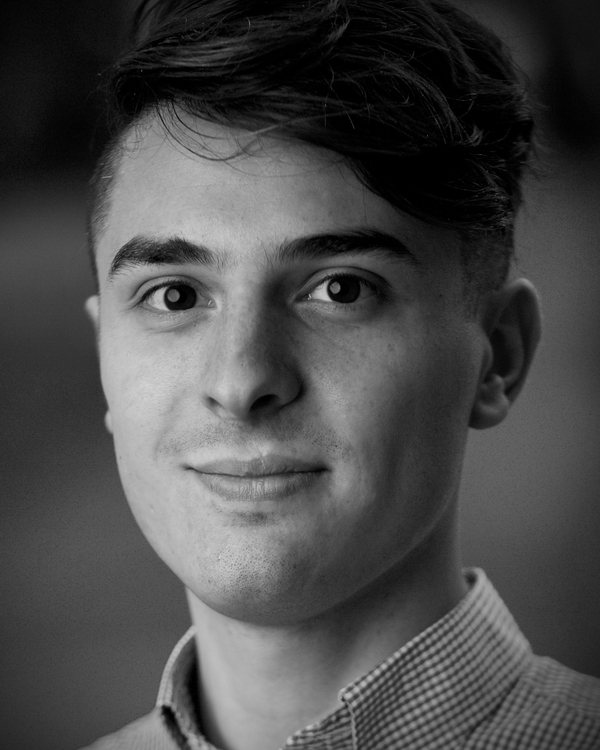}}]{Alex Cummaudo} is an \mbox{R\&D Software Engineer} and holds an \mbox{Associate Research Fellowship} at the \mbox{Applied Artificial Intelligence Institute} (A²I²), Deakin University, Australia. Alex received a BIT(Hons) in 2017 from Deakin, where he developed a novel computer vision pipeline to detect alphanumeric sequences in unstructured images, and a BSc in 2016 from Swinburne University of Technology, Australia. He is interested in abstracting pre-trained machine learning models through software components, and explores ways to enhance developer productivity and DevX of AI-based software. Alex is working towards completing his PhD at A²I², where he developed a novel software architecture tactic to improve integration reliability of AI-based components with traditional software, based on a thorough investigation of the behavioural and evolutionary profile of intelligent web services.
\end{IEEEbiography}
\begin{IEEEbiography}[{\includegraphics[width=1in,height=1.25in,clip,keepaspectratio]{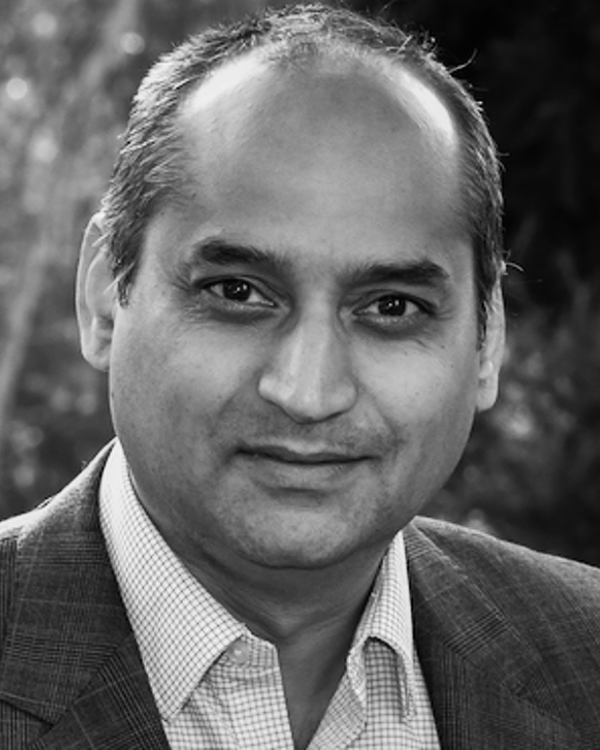}}]{Rajesh Vasa} is a professor of software and technology innovation at Deakin University and currently leads transnational research at the Applied Artificial Intelligence Institute. He has more than 2 decades of experience spanning both industry and academia with deep skills in data science, artificial intelligence and complex software systems design.  His career spans roles in development, operations and executive leadership in projects and organisations across the world.  Recent work included building intelligent homes for aged care, reducing traffic congestion, deep learning and neural networks in healthcare, and automating the process of innovation.  In the context of software engineering, his focus is on robust AI systems, sustainable software evolution, and software architecture.
\end{IEEEbiography}
\begin{IEEEbiography}[{\includegraphics[width=1in,height=1.25in,clip,keepaspectratio]{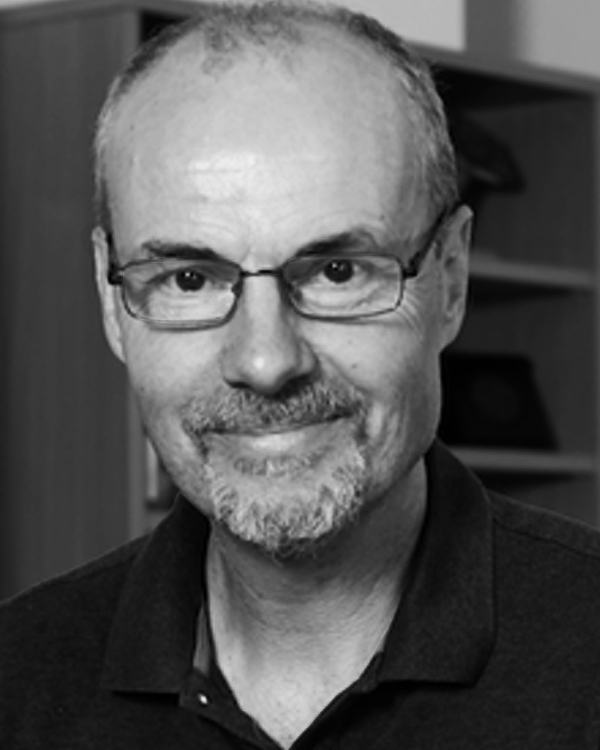}}]{John Grundy}
is Australian Laureate Fellow and Professor of Software Engineering at Monash University, Australia. He has published widely in automated software engineering, domain-specific visual languages, model-driven engineering, software architecture, and empirical software engineering, among many other areas. He is Fellow of Automated Software Engineering and Fellow of Engineers Australia.
\end{IEEEbiography}
\begin{IEEEbiography}[{\includegraphics[width=1in,height=1.25in,clip,keepaspectratio]{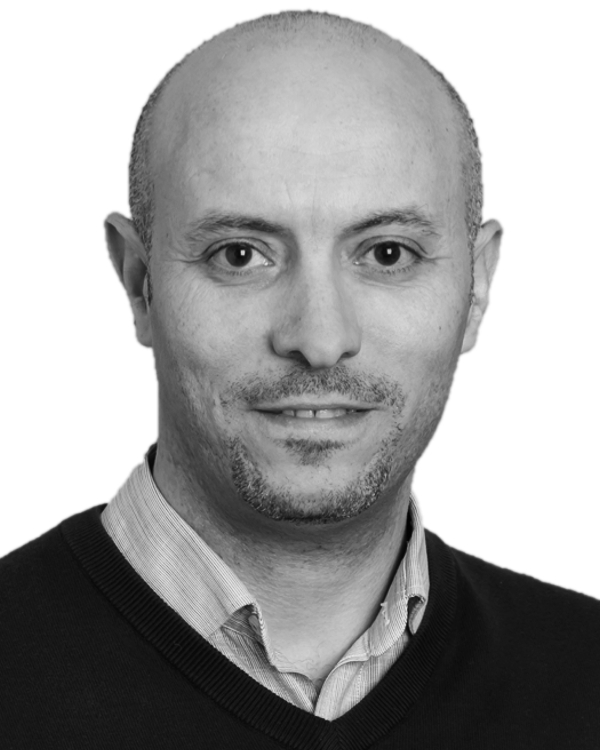}}]{Mohamed Abdelrazek}
is Associate Professor of Software Engineering and IoT. Mohamed has more than 15 years of the software industry, research and teaching experience. Before joining Deakin University in 2015, Mohamed worked as a senior research fellow at Swinburne University of Technology and Swinburne-NICTA software innovation lab (SSIL). Mohamed moved to Australia in 2010 to do his PhD in cloud computing security. Before that, Mohamed was the head of software development department at Microtech, a large software house, managing large-scale products/projects and managing large teams. Mohamed has deep experience in designing, developing, integrating, and managing large-scale software systems. Mohamed's research interests include Automated Software Engineering for Artificial Intelligence, Software Security, and Human-Centric Design.
\end{IEEEbiography}

\newpage\clearpage\onecolumn
\appendices
\pagenumbering{gobble}

\section{Detailed Overview of Our Proposed Taxonomy}
\label[appendix]{tse2020:tab:taxonomy}

Detailed descriptions of the 5 requirements of good API documentation (dimensions) and 34 generalised API documentation artefacts (categories/sub-dimensions) that help satisfy these requirements within our proposed taxonomy. Descriptions of examples of these documentation artefacts are italicised and provided for illustrative purposes. ILS = In-Literature Score, calculated as a ratio of papers that investigated or reported various issues concerning each artefact. IPS = In-Practice Score, calculated as the average response from our survey instrument. Colour scales indicate relevancy weight within ILS or IPS values for comparative purposes, where red = \textit{lowest} and green = \textit{highest}. GCV, AWS, ACV = Presence of category in Google Cloud Vision, Amazon Rekognition, and Azure Cloud Vision documentation. Presence indicated as \textit{fully present} (\circlepresent{}), \textit{partially present} (\circlepartialpresent{}), and \textit{not present} (\circlenotpresent{}).

{\def\cn{}
\def\cy{\checkmark}
\footnotesize
\begin{longtable}{rp{0.45\linewidth}|p{0.125\linewidth}|cc|ccc}
  \toprule
  \textbf{Key} &
  \textbf{Description} &
  \textbf{Primary Sources} &
  \textbf{ILS} &
  \textbf{IPS} & 
  \textbf{GCV} &
  \textbf{AWS} &
  \textbf{ACV} \\
  \midrule
  \midrule
  \endfirsthead

  \toprule
  \textbf{Key} &
  \textbf{Description} &
  \textbf{Primary Sources} &
  \textbf{ILS} &
  \textbf{IPS} &
  \textbf{GCV} &
  \textbf{AWS} &
  \textbf{ACV} \\
  \midrule
  \midrule
  \endhead
  \bottomrule
  
  \multicolumn{8}{r}{\textit{Continued on next page...}}\\
  \endfoot
  \bottomrule
  \endlastfoot

  \textbf{A}&
  \multicolumn{7}{l}{\textbf{Requirement 1: API Documentation should include Descriptions of API Usage}}\\
  \midrule

  A1&
  Quick-start guides; \textit{i.e., a guide to rapidly get started using the API in a specific programming language.}&
  \scriptsize S4, S9, S10 &
  \cellcolor[HTML]{f6b26b}Low&\cellcolor[HTML]{57bb8a}V High&\circlepresent{}&\circlepartialpresent{}&\circlepresent{}\\
  
  A2&
  Low-level reference manual; \textit{i.e., a manual documenting all API components to review fine-grade detail.}&
  \scriptsize S1, S3, S4, S8, S9, S10, S11, S12, S15, S16, S17 &
  \cellcolor[HTML]{a7c47d}High&\cellcolor[HTML]{a7c47d}High&\circlepresent{}&\circlepresent{}&\circlepresent{}\\
  
  A3&
  Explanation of high level architecture; \textit{i.e., explanations of the API's high-level architecture to better understand intent and context.}
  &
  \scriptsize S1, S2, S4, S11, S14, S16, S19, S20 &
  \cellcolor[HTML]{ffd666}Med&\cellcolor[HTML]{57bb8a}V High&\circlepresent{}&\circlepresent{}&\circlepresent{}\\

  A4&
  Introspection source code comments; \textit{i.e., code implementation and code comments (where applicable) to understand the API author's mindset.}
  &
  \scriptsize S1, S4, S7, S12, S13, S17, S20 &
  \cellcolor[HTML]{ffd666}Med&\cellcolor[HTML]{a7c47d}High&\circlenotpresent{}&\circlenotpresent{}&\circlenotpresent{}\\

  {A5}&
  Code snippets of basic component function; \textit{i.e., {code snippets (with comments) of no more than 30 LoC to understand a basic component functionality within the API.}}
  &
  \scriptsize {S1, S2, S4, S5, S6, S7, S9, S10, S11, S14, S15, S16, S18, S20, S21} &
  \cellcolor[HTML]{57bb8a}V High&\cellcolor[HTML]{57bb8a}V High&\circlepresent{}&\circlepresent{}&\circlepresent{}\\

  {A6}&
  Step-by-step tutorials with multiple components; \textit{i.e., {step-by-step tutorials, with screenshots to understand  how to build a non-trivial piece of functionality with multiple components of the API.}}
  &
  \scriptsize {S1, S2, S4, S5, S7, S9, S10, S15, S16, S18, S20, S21} &
  \cellcolor[HTML]{57bb8a}V High&\cellcolor[HTML]{57bb8a}V High&\circlepartialpresent{}&\circlepresent{}&\circlepresent{}\\

  A7&
  Downloadable production-ready source code; \textit{i.e., downloadable source code of production-ready applications that use the API to understand implementation in a large-scale solution.}
  &
  \scriptsize S1, S2, S5, S9, S15 &
  \cellcolor[HTML]{f6b26b}Low&\cellcolor[HTML]{57bb8a}V High&\circlepartialpresent{}&\circlepartialpresent{}&\circlepresent{}\\

  A8&
  Best-practices of implementation; \textit{i.e., best-practices of implementation to assist with debugging and efficient use of the API.}
  &
  \scriptsize S1, S2, S4, S5, S7, S8, S9, S14 &
  \cellcolor[HTML]{ffd666}Med&\cellcolor[HTML]{57bb8a}V High&\circlenotpresent{}&\circlepresent{}&\circlepartialpresent{}\\

  A9&
  An exhaustive list of all components; \textit{i.e., a list of all the major components that exist within the API.}
  &
  \scriptsize S4, S16, S19 &
  \cellcolor[HTML]{f6b26b}Low&\cellcolor[HTML]{57bb8a}V High&\circlenotpresent{}&\circlepresent{}&\circlepresent{}\\

  A10&
  Minimum system requirements to use the API; \textit{i.e., requirements and the dependencies to use the API on a particular system.}
  &
  \scriptsize S4, S7, S13, S17, S19 &
  \cellcolor[HTML]{f6b26b}Low&\cellcolor[HTML]{57bb8a}V High&\circlepartialpresent{}&\circlenotpresent{}&\circlepartialpresent{}\\
  
  A11&
  Instructions to install/update the API and its release cycle; \textit{i.e., instructions to install or begin using the API and details on its release cycle and how to update it.}
  &
  \scriptsize S4, S7, S8, S9, S11, S13, S16, S19 &
  \cellcolor[HTML]{ffd666}Med&\cellcolor[HTML]{57bb8a}V High&\circlepartialpresent{}&\circlepartialpresent{}&\circlenotpresent{}\\

  A12&
  Error definitions describing how to address problems
  &
  \scriptsize S1, S2, S4, S5, S9, S11, S13 &
  \cellcolor[HTML]{ffd666}Med&\cellcolor[HTML]{57bb8a}V High&\circlepartialpresent{}&\circlenotpresent{}&\circlenotpresent{}\\

  \midrule
  \textbf{B}&
  \multicolumn{7}{l}{\textbf{Requirement 2: API Documentation should include Descriptions of the API's Design Rationale}}\\
  \midrule
  
  {B1}&
  Entry-point purpose of the API; \textit{i.e., {a brief description of the purpose or overview of the API as a low barrier to entry.}}
  &
  \scriptsize {S1, S2, S4, S5, S6, S8, S10, S11, S15, S16} &
  \cellcolor[HTML]{a7c47d}High&\cellcolor[HTML]{57bb8a}V High&\circlepresent{}&\circlepresent{}&\circlepresent{}\\

  B2&
  What the API can develop; \textit{i.e., descriptions of concrete types of applications the API can develop.}
  &
  \scriptsize S2, S4, S9, S11, S15, S18 &
  \cellcolor[HTML]{ffd666}Med&\cellcolor[HTML]{57bb8a}V High&\circlepartialpresent{}&\circlepartialpresent{}&\circlepresent{}\\

  B3&
  Who should use the API; \textit{i.e., descriptions of the types of users who should use the API.}
  &
  \scriptsize S4, S9 &
  \cellcolor[HTML]{e67c73}V Low&\cellcolor[HTML]{a7c47d}High&\circlepartialpresent{}&\circlenotpresent{}&\circlenotpresent{}\\

  B4&
  Who will use the applications built using the API; \textit{i.e., descriptions of the types of users who will use the product the API creates.}
  &
  \scriptsize S4 &
      \cellcolor[HTML]{e67c73}V Low&\cellcolor[HTML]{ffd666}Med&\circlenotpresent{}&\circlenotpresent{}&\circlenotpresent{}\\

  B5&
  Success stories on the API; \textit{i.e., example success stories of major users that describe how well the API was used in production.}
  &
  \scriptsize S4 &
  \cellcolor[HTML]{e67c73}V Low&\cellcolor[HTML]{57bb8a}V High&\circlepartialpresent{}&\circlepresent{}&\circlepresent{}\\

  B6&
  Documentation comparing similar APIs to this API
  &
  \scriptsize S2, S6, S13, S18 &
  \cellcolor[HTML]{f6b26b}Low&\cellcolor[HTML]{a7c47d}High&\circlepartialpresent{}&\circlenotpresent{}&\circlepresent{}\\

  B7&
  Limitations on what the API can/cannot provide
  &
  \scriptsize S4, S5, S8, S9, S14, S16 &
  \cellcolor[HTML]{ffd666}Med&\cellcolor[HTML]{57bb8a}V High&\circlenotpresent{}&\circlepresent{}&\circlepresent{}\\

  \midrule
  \textbf{C}&
  \multicolumn{7}{l}{\textbf{Requirement 3: API Documentation should include Descriptions of the Domain Concepts behind the API}}\\
  \midrule
  
  C1&
  Relationship between API components and domain concepts
  &
  \scriptsize S3, S10 &
  \cellcolor[HTML]{e67c73}V Low&\cellcolor[HTML]{a7c47d}High&\circlenotpresent{}&\circlenotpresent{}&\circlepresent{}\\

  C2&
  Definitions of domain terminology; \textit{i.e., definitions of the domain-terminology and concepts, with synonyms if applicable.}
  &
  \scriptsize S2, S3, S4, S6, S7, S10, S14, S16 &
  \cellcolor[HTML]{ffd666}Med&\cellcolor[HTML]{57bb8a}V High&\circlepartialpresent{}&\circlenotpresent{}&\circlepartialpresent{}\\

  C3&
  Documentation for nontechnical audiences; \textit{i.e., generalised documentation for non-technical audiences regarding the API and its domain.}
  &
  \scriptsize S4, S8, S16 &
  \cellcolor[HTML]{f6b26b}Low&\cellcolor[HTML]{a7c47d}High&\circlepresent{}&\circlepresent{}&\circlepresent{}\\

  \midrule
  \textbf{D}&
  \multicolumn{7}{l}{\textbf{Requirement 4: API Documentation should include Additional Support Artefacts to aide Developer Productivity}}\\
  \midrule
  
  D1&
  FAQs
  &
  \scriptsize S4, S7 &
  \cellcolor[HTML]{e67c73}V Low&\cellcolor[HTML]{57bb8a}V High&\circlepresent{}&\circlepresent{}&\circlepresent{}\\

  D2&
  Troubleshooting hints
  &
  \scriptsize S4, S8 &
  \cellcolor[HTML]{e67c73}V Low&\cellcolor[HTML]{a7c47d}High&\circlenotpresent{}&\circlepartialpresent{}&\circlenotpresent{}\\

  D3&
  API diagrams; \textit{i.e., diagrammatically representing API components using visual architectural representations.}
  &
  \scriptsize S6, S13, S20 &
  \cellcolor[HTML]{f6b26b}Low&\cellcolor[HTML]{57bb8a}V High&\circlenotpresent{}&\circlenotpresent{}&\circlenotpresent{}\\

  D4&
  Contact for technical support
  &
  \scriptsize S4, S8, S19 &
  \cellcolor[HTML]{f6b26b}Low&\cellcolor[HTML]{ffd666}Med&\circlepresent{}&\circlepresent{}&\circlepresent{}\\

  D5&
  Printed guide
  &
  \scriptsize S4, S6, S7, S9, S16 &
  \cellcolor[HTML]{f6b26b}Low&\cellcolor[HTML]{57bb8a}V High&\circlenotpresent{}&\circlepresent{}&\circlepresent{}\\

  D6&
  Licensing information
  &
  \scriptsize S7 &
  \cellcolor[HTML]{e67c73}V Low&\cellcolor[HTML]{57bb8a}V High&\circlenotpresent{}&\circlenotpresent{}&\circlepartialpresent{}\\

  \midrule
  \textbf{E}&
  \multicolumn{7}{l}{\textbf{Requirement 5: API Documentation should be Presented in an Easily Digestible Format}}\\
  \midrule
  
  E1&
  Searchable knowledge base
  &
  \scriptsize S3, S4, S6, S10, S14, S17, S18 &
  \cellcolor[HTML]{ffd666}Med&\cellcolor[HTML]{57bb8a}V High&\circlepresent{}&\circlepresent{}&\circlepresent{}\\

  E2&
  Context-specific discussion forums
  &
  \scriptsize S4, S10, S11 &
  \cellcolor[HTML]{f6b26b}Low&\cellcolor[HTML]{57bb8a}V High&\circlepresent{}&\circlepresent{}&\circlepartialpresent{}\\

  E3&
  Quick-links to other relevant components
  &
  \scriptsize S6, S16, S20 &
  \cellcolor[HTML]{f6b26b}Low&\cellcolor[HTML]{57bb8a}V High&\circlenotpresent{}&\circlenotpresent{}&\circlenotpresent{}\\

  E4&
  Structured navigation style; \textit{i.e., breadcrumbs}
  &
  \scriptsize S6, S10, S20 &
  \cellcolor[HTML]{f6b26b}Low&\cellcolor[HTML]{a7c47d}High&\circlepresent{}&\circlepresent{}&\circlepresent{}\\

  E5&
  Visualised map of navigational paths; \textit{i.e., to certain API components in the website.}
  &
  \scriptsize S6, S14, S20 &
  \cellcolor[HTML]{f6b26b}Low&\cellcolor[HTML]{57bb8a}V High&\circlenotpresent{}&\circlenotpresent{}&\circlenotpresent{}\\

  {E6}&
  Consistent look and feel
  &
  \scriptsize {S1, S2, S3, S5, S6, S8, S10, S15, S20} &
  \cellcolor[HTML]{a7c47d}High&\cellcolor[HTML]{57bb8a}V High&\circlepresent{}&\circlepresent{}&\circlepresent{}\\
\end{longtable}
\normalsize}

\clearpage
\section{Sources of Documentation}\label[appendix]{tse2020:tab:docsources}
Sources of documentation used for the validation of the taxonomy. For clarity, exact webpages are not referenced for each category, but can be found in supplementary materials.
\small
\begin{longtable}{p{.17\linewidth}|p{.75\linewidth}}
  \toprule
  \textbf{Service} & \textbf{Document Sources}\\
  \midrule
  \endfirsthead
  \toprule
  \textbf{Service} & \textbf{Document Sources}\\
  \midrule
  \endhead
  \bottomrule
  \multicolumn{2}{r}{\textit{Continued on next page...}}\\
  \endfoot
  \bottomrule
  \endlastfoot
    Google Cloud Vision &
    \vspace{-1.75mm}
    \begin{itemize}[label=,leftmargin=10pt,topsep=0pt,partopsep=0pt,noitemsep,nolistsep,itemindent=-10pt]
\item \url{https://cloud.google.com/vision/docs/quickstart-client-libraries}
\item \url{https://googleapis.github.io/google-cloud-java/google-cloud-clients/apidocs/index.html}
\item \url{https://cloud.google.com/vision/#cloud-vision-use-cases}
\item \url{https://cloud.google.com/vision/docs/quickstart-client-libraries#using_the_client_library}
\item \url{https://cloud.google.com/vision/docs/tutorials}
\item \url{https://cloud.google.com/community/tutorials?q=vision}
\item \url{https://cloud.google.com/vision/docs/samples#mobile_platform_examples}
\item \url{https://cloud.google.com/docs/enterprise/best-practices-for-enterprise-organizations}
\item \url{https://cloud.google.com/functions/docs/bestpractices/tips}
\item \url{https://cloud.google.com/vision/#derive-insight-from-images-with-our-powerful-cloud-vision-api}
\item \url{https://cloud.google.com/vision/docs/quickstart-client-libraries}
\item \url{https://cloud.google.com/vision/docs/release-notes}
\item \url{https://cloud.google.com/vision/docs/reference/rpc/google.rpc#google.rpc.Code}
\item \url{https://cloud.google.com/vision/#derive-insight-from-your-images-with-our-powerful----------pretrained-api-models-or-easily-train-custom-vision-models-with-automl----------vision-beta}
\item \url{https://cloud.google.com/vision/#insight-from-your-images}
\item \url{https://developers.google.com/machine-learning/glossary/}
\item \url{https://cloud.google.com/vision/docs/resources}
\item \url{https://cloud.google.com/vision/sla}
\item \url{https://cloud.google.com/vision/docs/data-usage}
\item \url{https://cloud.google.com/vision/docs/support#searchbox}
\item \url{https://cloud.google.com/vision/docs/support}
    \end{itemize}\\
    Amazon Rekgonition &
    \vspace{-1.75mm}
    \begin{itemize}[label=,leftmargin=10pt,topsep=0pt,partopsep=0pt,noitemsep,nolistsep,itemindent=-10pt]
\item \url{https://docs.aws.amazon.com/rekognition/latest/dg/getting-started.html}
\item \url{https://docs.aws.amazon.com/AWSJavaSDK/latest/javadoc/index.html}
\item \url{https://aws.amazon.com/blogs/machine-learning/using-amazon-rekognition-to-identify-persons-of-interest-for-law-enforcement/}
\item \url{https://aws.amazon.com/rekognition/#Rekognition_Image_Use_Cases}
\item \url{https://docs.aws.amazon.com/rekognition/latest/dg/labels-detect-labels-image.html}
\item \url{https://aws.amazon.com/rekognition/getting-started/#Tutorials}
\item \url{https://aws.amazon.com/blogs/machine-learning/category/artificial-intelligence/amazon-rekognition/}
\item \url{https://docs.aws.amazon.com/code-samples/latest/catalog/code-catalog-java-example_code-rekognition.html}
\item \url{https://docs.aws.amazon.com/rekognition/latest/dg/best-practices.html}
\item \url{https://docs.aws.amazon.com/rekognition/latest/dg/API_Operations.html}
\item \url{https://aws.amazon.com/rekognition/image-features/}
\item \url{https://aws.amazon.com/releasenotes/?tag=releasenotes%23keywords%23amazon-rekognition}
\item \url{https://docs.aws.amazon.com/rekognition/latest/dg/setting-up.html}
\item \url{https://aws.amazon.com/rekognition/}
\item \url{https://aws.amazon.com/rekognition/}
\item \url{https://docs.aws.amazon.com/rekognition/latest/dg/limits.html}
\item \url{https://aws.amazon.com/rekognition/pricing/}
\item \url{https://aws.amazon.com/rekognition/sla/}
\item \url{https://aws.amazon.com/rekognition/faqs/}
\item \url{https://docs.aws.amazon.com/rekognition/latest/dg/video-troubleshooting.html}
\item \url{https://docs.aws.amazon.com/rekognition/latest/dg/rekognition-dg.pdf}
\item \url{https://github.com/awsdocs/amazon-rekognition-developer-guide/issues}
\item \url{https://forums.aws.amazon.com/thread.jspa?threadID=285910}
    \end{itemize}\\
    Azure Computer Vision &
    \vspace{-1.75mm}
    \begin{itemize}[label=,leftmargin=10pt,topsep=0pt,partopsep=0pt,noitemsep,nolistsep,itemindent=-10pt]
\item \url{https://docs.microsoft.com/en-au/azure/cognitive-services/computer-vision/quickstarts-sdk/csharp-analyze-sdk}
\item \url{https://docs.microsoft.com/en-us/java/api/overview/azure/cognitiveservices/client/computervision?view=azure-java-stable}
\item \url{https://docs.microsoft.com/en-us/azure/architecture/example-scenario/ai/intelligent-apps-image-processing}
\item \url{https://docs.microsoft.com/en-us/azure/cognitive-services/computer-vision/tutorials/java-tutorial}
\item \url{https://docs.microsoft.com/en-us/azure/cognitive-services/custom-vision-service/logo-detector-mobile}
\item \url{https://docs.microsoft.com/en-au/azure/cognitive-services/computer-vision/tutorials/storage-lab-tutorial}
\item \url{https://docs.microsoft.com/en-us/azure/cognitive-services/computer-vision/tutorials/csharptutorial}
\item \url{https://docs.microsoft.com/en-us/azure/cognitive-services/custom-vision-service/getting-started-improving-your-classifier}
\item \url{https://docs.microsoft.com/en-au/azure/cognitive-services/computer-vision/home#analyze-images-for-insight}
\item \url{https://docs.microsoft.com/en-au/azure/cognitive-services/computer-vision/vision-api-how-to-topics/howtocallvisionapi}
\item \url{https://docs.microsoft.com/en-us/azure/cognitive-services/custom-vision-service/release-notes}
\item \url{https://docs.microsoft.com/en-au/azure/cognitive-services/computer-vision/}
\item \url{https://azure.microsoft.com/en-au/services/cognitive-services/computer-vision/}
\item \url{https://azure.microsoft.com/en-us/pricing/details/cognitive-services/computer-vision/}
\item \url{https://docs.microsoft.com/en-au/azure/cognitive-services/computer-vision/concept-tagging-images}
\item \url{https://docs.microsoft.com/en-au/azure/cognitive-services/computer-vision/home}
\item \url{https://azure.microsoft.com/en-us/support/legal/sla/cognitive-services/v1_1/}
\item \url{https://docs.microsoft.com/en-au/azure/cognitive-services/computer-vision/faq}
\item \url{https://azure.microsoft.com/en-us/support/legal/}
    \end{itemize}\\
\end{longtable}
\normalsize

\clearpage
\section{List of Online Artefacts}\label[appendix]{tse2020:sec:online-artefacts}
\bibliographystyleW{model1-num-names}
\bibliographyW{webservices,webdocimprovements}

\newpage\clearpage
\section{List of Primary Sources}\label[appendix]{tse2020:sec:primary-sources}
\bibliographystyleS{model1-num-names}

List of the primary sources found from our systematic mapping study. Each citation is referenced by a prefixed `S'. We also list the respective citation count, as measured by the number of citations the publication has from Google Scholar as at July 2020. We also list the venue ranking (as at 2020), as measured by Scimago Rankings or Qualis Ranking for Journals and CORE Rankings for conference publications. If no rank can be found, a dash is used.

\nobibliography*

\small
\begin{longtable}{p{0.03\linewidth}p{.79\linewidth}|cc}
  \toprule
  \textbf{Ref} & \textbf{Citation} & \textbf{Cite\#} & \textbf{Rank}\\
  \midrule
  \endfirsthead
  \toprule
  \textbf{Ref} & \textbf{Citation} & \textbf{Cite\#} & \textbf{Rank}\\
  \midrule
  \endhead
  \bottomrule
  \multicolumn{4}{r}{\textit{Continued on next page...}}\\
  \endfoot
  \bottomrule
  \endlastfoot
  
\relax[S1]&\bibentry{Robillard:2009uk}&305&Q1\\
\relax[S2]&\bibentry{Robillard:2011uv}&254&Q1\\
\relax[S3]&\bibentry{Ko:2011fb}&33&A\\
\relax[S4]&\bibentry{Nykaza:2002td}&56&--\\
\relax[S5]&\bibentry{Watson:2013fx}&14&B1\\
\relax[S6]&\bibentry{Jeong:2009tu}&34&--\\
\relax[S7]&\bibentry{Aghajani:2019bo}&6&A*\\
\relax[S8]&\bibentry{Haselbock:2018jd}&2&C\\
\relax[S9]&\bibentry{Inzunza:2018dn}&3&C\\
\relax[S10]&\bibentry{Meng:2017cx}&12&Q1\\
\relax[S11]&\bibentry{Geiger:2018fv}&4&Q1\\
\relax[S12]&\bibentry{Head:2018baa}&4&A*\\
\relax[S13]&\bibentry{Aversano:2017ic}&4&--\\
\relax[S14]&\bibentry{Robillard:hk}&55&A*\\
\relax[S15]&\bibentry{Watson:2012uy}&10&B1\\
\relax[S16]&\bibentry{Maalej2013}&110&Q1\\
\relax[S17]&\bibentry{Parnas:2007fb}&2&B\\
\relax[S18]&\bibentry{Bottomley:2005fs}&0&--\\
\relax[S19]&\bibentry{Taulavuori:2004el}&40&Q1\\
\relax[S20]&\bibentry{Kotula:1998wp}&27&Q1\\
\relax[S21]&\bibentry{McLellan:1998vu}&105&Q1\\

\end{longtable}
\normalsize

\clearpage
\section{Detailed Suggested Improvements}\label[appendix]{tse2020:sec:suggested-improvements}

For this assessment, we select the ILS or IPS values for categories that are considered either somewhat or very helpful (i.e., a score greater than 0.50). We then match these against categories that are found to be partially or not present within each service. In total, we found 12 categories where improvements can be made across all dimensions except \dime{}, detailed below .

\subsection[Dimension A Issues]{Issues regarding \dima{}}

\noindent
\textbf{Quick-start guides \dimcat{A1}:} 
Quick-start guides should provide a short tutorial that allows programmers to pick up the basics of an API in a programming language of their choice. For the services assessed, each offer various client SDKs (e.g., as Java or Python client libraries). Google Cloud Vision and Azure Computer Vision offer quick-start guides \citepweb{Quicksta92:online,Quicksta70:online} in which sets of articles target various SDKs or are client-agnostic with code snippets that can be changed to the client language/SDK of the developer's choice. Amazon Rekognition offers exercises in setting up the AWS SDK and using the command-line interface to interact with image analysis components \citepweb{Exercise86:online}, however this is client-agnostic nor does it provide details in how to get started with using the client SDKs.

\def\SuggestedImprovement{\noindent\itshape\small\textbf{\faHandORight{} Suggested improvement:~}}

\begin{leftbar}
\SuggestedImprovement
Ensure tutorials detail \uline{all} client-libraries and how developers can produce a minimum working example using the service on their own computer using that client library. For each SDK offered, there should be details on how to install, authenticate and use a component using local data. For example, this may be as simple as using the service to determine if an image of a dog contains the label `dog'.
\end{leftbar}

\noindent
\textbf{Step-by-step tutorials \dimcat{A6}:}  Google Cloud Vision offers tutorials limited to one component. These do not sufficiently demonstrate how to combine \textit{multiple components} of the API together and how developers should integrate it with a different platform, which a good step-by-step tutorial should detail. The official AWS Machine Learning blog \citepweb{AmazonRe55:online} provides extensive tutorials (in some cases, with a suggested tutorial completion time of over an hour) that integrate multiple Amazon Rekognition components with other AWS components. Microsoft provide tutorials \citepweb{Tutorial42:online,Tutorial6:online,GitHubAz56:online} integrating multiple components within their service to mobile applications and the Azure platform. 

\begin{leftbar}
\SuggestedImprovement
Ensure tutorials combine \uline{multiple} components of the service together, are extensive, and require developers to spend a non-trivial amount of time to produce a basic application. For example, the tutorial may detail how to integrate the API into a smartphone application to achieve the following: (i) take a photo with the camera, (ii) detect if a person is within the image, (iii) analyse the visual features of the person.
\end{leftbar}

\noindent
\textbf{Downloadable production-ready applications \dimcat{A7}:} Microsoft provide a downloadable application \citepweb{SampleEx87:online} that explores many components of the Azure Computer Vision API. The application is thoroughly documented with and also provides guidance on how to structure the architecture design of the program. While Rekognition and Google Cloud Vision also provide downloadable source code, they are largely under-documented, do not combine multiple components of the API together, and only use god-classes to handle all requests to the API \citepweb{JavaSDKV25:online,SampleAp41:online}.

\begin{leftbar}
\SuggestedImprovement
Downloadable source code should be thoroughly documented, and should avoid the use of god-classes that demonstrate a single piece of the service's functionality. Ideally, the \uline{architecture} of a production-ready application should be demonstrated to developers.
\end{leftbar}

\noindent
\textbf{Understanding best-practices \dimcat{A8}:} Google Cloud provides best-practices for its platform in both general and enterprise contexts \citepweb{Bestprac23:online,TipsTric26:online}, but there is little advice provided to guide developers on how best to use Google Cloud \textit{Vision}. Microsoft provides guidance on improving results of custom vision classifiers \citepweb{Improvin23:online}, but no further details on non-custom vision classifiers are found. We found the most detailed best-practices to be provided by Amazon Rekognition \citepweb{BestPrac58:online}, which outlines more detailed strategies such as reducing data transfer by storing and referencing images on S3 Buckets or the attributes images should have in various scenarios (e.g., the angles of a person's face in facial recognition).

\begin{leftbar}
\SuggestedImprovement
Document best-practices for all major components of the computer vision service. Guide developers on the types of input data that produce the best results, advisable minimum image sizes and recommended file types, and suggest ways to overcome limitations that improve usage and cost efficiency. Provide guidance in more than one use case; give a range of scenarios that demonstrate different best practices for different domains.
\end{leftbar}

\noindent
\textbf{Exhaustive lists of all major API components \dimcat{A9}:} Amazon provides a two-fold feature list that describes both the key features of Rekognition at a high-level \citepweb{AmazonRe70:online} as well as a detailed, technical breakdown of each API operation provided within the service \citepweb{ActionsA39:online}. Microsoft also provide a list of high-level features that Azure Computer Vision can analyse \citepweb{WhatisCo90:online} which provides hyperlinks to detailed descriptions of each feature. Google's Cloud Vision API provides a partial breakdown of the types of services provided, however this list is not fully complete, nor are there hyperlinks to more detailed descriptions of each of the features \citepweb{VisionAI32:online}.

\begin{leftbar}
\SuggestedImprovement
Document key features that the computer vision classifier can perform at a high level. This should be easy to find from the service's landing page. Each feature should be described with reference to more detailed descriptions of the feature's exact API endpoint and required inputs, outputs and possible errors.
\end{leftbar}

\noindent
\textbf{Minimum system requirements and dependencies \dimcat{A10}:} Although there is no dedicated webpage for this on any of the services investigated, there are listed dependencies for the client libraries in Google's and Azure's quick-start guides \citepweb{Quicksta92:online,CalltheC0:online}. These may be embedded within the quick-start guide as developers are likely to encounter dependency issues when they first start using the API. We found it a challenge to discover similar documentation this in Amazon's documentation.

\begin{leftbar}
\SuggestedImprovement
Any system requirements and dependency issues should be well-highlighted within the documentation's quick-start guide; developers are likely to encounter these issues within the early stages of using an API, and it is highly relevant to provide solutions to these issues within the quick-starts.
\end{leftbar}

\noindent
\textbf{Installation and release cycle notes \dimcat{A11}:} It is imperative that developers know what has changed between releases and how frequently the releases are exported. We found release notes for Amazon Computer Vision, although they are only major releases and have not been updated since 2017 \citepweb{AWSRelea46:online} which does not account for evolution in the service's responses \citep{Cummaudo:2019icsme}. Google's and Microsoft's release notes are generally more frequently updated, therefore developers can get a sense of its release frequency \citepweb{ReleaseN91:online,ReleaseN4:online}. However, there are evolution issues that are not addressed. Installation instructions are detailed within Rekognition's developer guide, outlining how to sign up for an account, and install the AWS command-line interface \citepweb{Step1Set76:online}.

\begin{leftbar}
\SuggestedImprovement
Ensure release notes detail label evolution, including any new additional labels that may have been introduced within the service. Transparency around the changes made to the service should go beyond new features: document potential changes that may influence maintenance of a system using the computer vision service so that developers are aware of potential side-effects of upgrading to a newer release.
\end{leftbar}

\subsection[Dimension B Issues]{Issues regarding \dimb{}}

\noindent
\textbf{Limitations of the API \dimcat{B7}:} The most detailed limitations documented were found on Rekognition's dedicated limitations page \citepweb{Limitsin66:online} that outlines functional limitations such as the maximum number of faces or words that can be detected in an image, the size requirements of images, and file type information. For the other services, functional limitations are generally found within each endpoint's API documentation, instead of within a dedicated page.

\begin{leftbar}\SuggestedImprovement
Document all functional limitations in a dedicated page that outline the maximum and minimum input requirements the classifier can handle. Documentation of the types of labels the service can provide is also desired.  
\end{leftbar}

\subsection[Dimension C Issues]{Issues regarding \dimc{}}

\noindent
\textbf{Conceptual understanding of the API \dimcat{C1}:} Azure Computer Vision provides `concept' pages describing the high-level concepts behind computer vision and where these functions are implemented within the APIs (e.g., \citepweb{Contentt49:online}). We were unable to find similar conceptual documentation for the other services assessed.

\begin{leftbar}\SuggestedImprovement
Document the concepts behind computer vision; differentiate between foundational concepts such as object localisation, object recognition, facial localisation and facial analysis such that developers are able to make the distinction between them. Relate these concepts back to the API and provide references to where the APIs implement these concepts.
\end{leftbar}

\noindent
\textbf{Definitions of domain-specific terminology \dimcat{C2}:}  Terminologies relevant to machine learning concepts powering these computer vision services are well detailed within Google's machine learning glossary \citepweb{MachineL36:online}, however few examples matching computer vision are immediately relevant. While this page is linked from the original Google Cloud Vision documentation, it may be too technical for application developers to grasp. A slightly better example of this is \citepweb{WhatisCo90:online}, where developers can understand computer vision terms in lay terms.

\begin{leftbar}\SuggestedImprovement
Current computer vision services use a myriad of terminologies to refer to the same conceptual feature; for example, while Microsoft refers to object recognition as `image tagging', Google refers to this as `label detection'. If a consolidation of terms is not possible, then computer vision services should provide a glossary that provides synonyms for these terminologies so that developers can easily move between service providers without needing to relink terms back to concepts.
\end{leftbar}

\subsection[Dimension D Issues]{Issues regarding \dimd{}}

\textbf{Troubleshooting suggestions \dimcat{D2}:} The only troubleshooting tips found in our analysis were in Rekognition's video service \citepweb{Troubles2:online}. Further detailed instances of these troubleshooting tips could be expanded to non-video issues. For instance, if developers upload `noisy' images, how can they inform the system of a specific ontology to use or to focus on parts of the foreground of background of the image? These are suggestions which we have proposed in prior work \citep{Cummaudo:2019icsme} that do not seem to be documented.

\begin{leftbar}\SuggestedImprovement
Ensure troubleshooting tips provide advice for testing against different types of valid input images.   
\end{leftbar}

\noindent
\textbf{Diagrammatic overview of the API \dimcat{D3}:} None of the computer vision services provide any overview of the API in terms of the features and processing steps on how they should be use. For instance, pre-processing and post-processing of input and response data should be considered and an understanding of how this fits into the `flow' of an application highlighted. Moreover, no UML diagrams could be found.

\begin{leftbar}\SuggestedImprovement
Provide diagrams illustrating the service within context of use, such as how it can be integrated with other service features or how a specific API endpoint may be used within a client application. Consider integrating interactive UML diagrams so that developers can easily explore various aspects of the documentation in a visual perspective.
\end{leftbar}

\clearpage
\section{Survey Questions}\label[appendix]{tse2020:sec:survey}

\def\AgreementScale{{\footnotesize \textit{[Strongly agree, Somewhat agree, Neither agree nor disagree, Somewhat disagree, Strongly disagree]}\bigskip}}

\noindent
This appendix contains the exact text of the survey described in \cref{tse2020:sec:validation:survey}. Our instrument also included questions where answers were not included in the research reported in this article, e.g. questions 1 and 2 regarding consent and ensuring participants have had development experience. Images used within the survey have been removed.

\bigskip
\hrule\sffamily\small

\subsection*{Developer opinions towards the importance of web API documentation recommendations}\noindent
In this study, we are finding out how important recommendations of web API documentation are to developers. From this, we will improve AI-powered APIs. While there are screenshots of example APIs in the questions, think of an API that you have used based on \textbf{your own prior experience} when answering these questions.   Thanks for taking the time to answer these questions; it should only take you about \textbf{10--20 minutes} to complete. 

\subsubsection*{Attribution Notice}\noindent
Portions of this questionnaire are reproduced from work created and shared by Google and used according to terms described in the Creative Commons 3.0 Attribution License. 

\bigskip\hrule

\subsubsection*{Implementation-specific documentation of web APIs}\noindent
When answering these questions please answer with respect to \textbf{your own experience} in learning web APIs (if applicable). Any examples provided exist solely to help illustrate the statement. For each question, please nominate how much you agree with the following statements: \AgreementScale

\begin{enumerate}[label=Q3\alph*.,leftmargin=2\parindent]
\item I think quick-start guides with code that help me get started with an API’s client library are important.
 e.g., quick-start guides that show how to get started and interact with the API and its responses.
\item I don't find low-level documentation of all classes and methods particularly helpful.
 e.g., a generated online reference manual from Javadoc comments.  
\item I would imagine that explanations of the API's high-level architecture, context and rationale would be important to better understand how to consume the API.
  e.g., a graphic showing how the API could fit into the wider context of an application.  
\item If I want to understand why an API did something that I didn't expect, the source code comments generally don't help me.
  e.g., an example from the Lodash API that describes why set.add isn't directly returned.  
\item I find small code snippets with comments to demonstrate a single component's basic functionality within the API a useful way to learn.
  e.g., 10-30 lines of code to demonstrating various how-tos of a computer vision API.  
\item  I think it's cumbersome to read through step-by-step tutorials that show how to build something non-trivial with multiple components using the API. 
   e.g., a ten-step tutorial documenting how to combine face recognition, face analysis, scene description, and landmark detection API components to generate descriptions of photos.   
\item  I think it's useful to download source code of production-ready applications that demonstrate the use of multiple facets of the API. 
   e.g., a downloadable iOS app that demonstrates how to perform image analysis on an iPhone/iPad.
\item  I think official documentation describing the ‘best-practices’ of how to use the API to assist with debugging and efficiency is not helpful. 
   e.g., an article describing the correct ways of doing things, the best tools to use, and how to write well-performing code.   
\item  I believe an exhaustive list of all major components in the API without excessive detail would be useful when learning an API. 
   e.g., a computer vision web API might list object detection, object localisation, facial recognition, and facial comparison as its 4 components.   
\item  I believe minimum system requirements and/or dependencies to use the API do not always need to be part of official documentation. 
   e.g., I can find descriptions of how to get started with a Python environment for a cloud platform on community forums instead of the API's website.   
\item  I think instructions on how to install or access the API, update it, and the frequency of its release cycle is all useful information to know about. 
   e.g., a list showing the latest releases, what was added and how to update your application to make use of it.   
\item  Error codes describing specific problems with an API are not helpful. 
   e.g., a list of canonical HTTP error codes and how to interpret them.   
\end{enumerate}

\bigskip\hrule
\subsubsection*{Rationale-specific documentation of web APIs}\noindent
When answering these questions please answer with respect to \textbf{your own experience} in learning web APIs (if applicable). Any examples provided exist solely to help illustrate the statement. For each question, please nominate how much you agree with the following statements: \AgreementScale

\begin{enumerate}[label=Q4\alph*.,leftmargin=2\parindent]
\item I think that, as a starting point when beginning to learn about an API, I would like to read about descriptions of the API's purpose and overview. 
\item I don't find descriptions of the types of applications the API can develop helpful. 
\item I believe that descriptions of the types of developers who should and shouldn't use the API is important to know. 
\item I don't think that descriptions of the types of end-users who will use the product built using the API is important to know in advance. 
\item I think that if I read success stories about when the API was previously used in production, I would have a better indicator of how I could use that API. 
\item I think that documentation that compares an API to other, similar APIs confusing and not important. 
\item I believe it is important to know about what the limitations are on what the API can and cannot provide. 
\end{enumerate}

\subsubsection*{Conceptual-specific documentation of web APIs}\noindent
When answering these questions please answer with respect to \textbf{your own experience} in learning web APIs (if applicable). Any examples provided exist solely to help illustrate the statement. For each question, please nominate how much you agree with the following statements: \AgreementScale

\begin{enumerate}[label=Q5\alph*.,leftmargin=2\parindent]
\item I wouldn’t read through theory about the API's domain that relates theoretical concepts to API components and how both work together. 
\item I think it is important to know the definitions of the API’s domain-specific terminology and concepts (with synonyms where needed). 
   e.g., a computer vision API that uses machine learning should list machine learning concepts. 
\item It's not really important to document information about the API to non-technical audiences, such as managers and other stakeholders. 
   e.g., pricing information, uptime information, QoS metrics/SLAs etc.   
\end{enumerate}

\bigskip\hrule
\subsubsection*{General-support documentation of web APIs}\noindent
When answering these questions please answer with respect to \textbf{your own experience} in learning web APIs (if applicable). Any examples provided exist solely to help illustrate the statement. For each question, please nominate how much you agree with the following statements: \AgreementScale

\begin{enumerate}[label=Q6\alph*.,leftmargin=2\parindent]
\item  I find lists of Frequently Asked Questions (FAQs) helpful. 
\item  When something goes wrong, I don't read through troubleshooting suggestions for specific problems straight away as I like to solve it myself. 
\item  I like to see diagrammatic representations of an API's components using visual architectural visualisations. 
   e.g., UML class diagram, sequence diagram. 
\item  I wouldn't look for email addresses and/or phone number for technical support in an API's documentation. 
\item  I generally refer to a programmer's reference guide or textbook about the API when I need to. 
\item  I don't think it's important to read about the licensing information about the API. 
\end{enumerate}

\bigskip\hrule
\subsubsection*{The effect of structure and tooling on web API documentation}\noindent
When answering these questions please answer with respect to \textbf{your own experience} in learning web APIs (if applicable). Any examples provided exist solely to help illustrate the statement. For each question, please nominate how much you agree with the following statements: \AgreementScale

\begin{enumerate}[label=Q7\alph*.,leftmargin=2\parindent]
\item I would like to use a searchable knowledge base to find information.
\item I think a context-specific discussion forum between developers isn't very helpful as it just introduces noise.  
  e.g., issue trackers, Slack group. 
\item I think links to other similar documentation frequently viewed by other developers would be useful. 
   e.g., 'people who viewed this also viewed…' 
\item If I get lost within the API's documentation, a 'breadcrumbs'-style of navigation isn't very useful to me. 
\item A visualised map of navigational paths to common API components in the website would be useful to have. 
   e.g., a large and complex API for Enterprise Service-Oriented Architecture where I could click into various boxes to read about components and arrows to read about how they are related.   
\item I believe ensuring consistent look and feel of all documentation isn't necessary to a good API documentation. 
\end{enumerate}

\bigskip\hrule
\subsubsection*{Demographics}\noindent

\begin{enumerate}[label=Q8\alph*.,leftmargin=2\parindent]
  \item Are you, or do you aspire to be, a professional programmer? Or would you consider programming a hobby?\\\noindent \textit{\footnotesize[Professional, Hobbyist]}
  \item How many years have you been programming? \\\noindent\textit{\footnotesize
[1--5 years,
6--10 years,
11--15 years,
16--20 years,
21--30 years,
31--40 years,
41+ years]}
  \item In what type of role would you say your current job falls into? \\\noindent\textit{\footnotesize
[
Back-end developer,
Data or business analyst,
Data scientist or machine learning specialist,
Database administrator,
Designer,
Desktop or enterprise applications developer,
DevOps specialist,
Educator or academic researcher,
Embedded applications or devices developer,
Engineering manager,
Front-end developer,
Full-stack developer,
Game or graphics developer,
Marketing or sales professional,
Mobile developer,
Product manager,
QA or test developer,
Student,
System administration]}
  \item What level of seniority would you say this role falls into? \\\noindent\textit{\footnotesize
[Intern Role,
Graduate Role,
Junior Role,
Mid-Tier Role,
Senior Role,
Lead Role,
Principal Role,
Management,
N/A (e.g., I am a student),
Other]}

  \item What industry would you say you work in? \\\noindent\textit{\footnotesize
[Cloud-based solutions or services,
Consulting,
Data and analytics,
Financial technology or services,
Healthcare technology or services,
Information technology,
Media, advertising, publishing, or entertainment,
Other software development,
Retail or eCommerce,
Software as a service (SaaS) development,
Web development or design,
N/A (e.g., I am a student),
Other industry not listed here]}
\end{enumerate}
\bigskip\hrule\bigskip
\hspace{\fill}\textit{** End of Survey **}\hspace{\fill}

\end{document}